\documentclass[sn-mathphys]{sn-jnl}% Math and Physical Sciences Reference Style
\usepackage{siunitx}
\usepackage[version=4]{mhchem}
\usepackage{multirow}

\usepackage{natbib}
\setcitestyle{super,open={},close={}}

\usepackage{soul}

\renewcommand{\thesection}{}% Remove section references...
\renewcommand{\thesubsection}{\arabic{subsection}}%... from subsections

\jyear{2024}%

\newcommand{\kt}{k_{\text{B}}T}
\def \deg {^{\circ}}

\newcommand{\tcr}{\textcolor{black}}

\begin{document}

\title[DNA Calorimetric Force Spectroscopy at Single Base Pair Resolution]{DNA Calorimetric Force Spectroscopy at Single Base Pair Resolution}


\author[1]{\fnm{P.} \sur{Rissone}}\email{rissone.paolo@gmail.com}

\author[2,3]{\fnm{M.} \sur{Rico-Pasto}}\email{m.rico.pasto@gmail.com}

\author[4]{\fnm{S. B.} \sur{Smith}}\email{stevesmith@gmail.com}

\author*[1,5]{\fnm{F.} \sur{Ritort}}\email{ritort@ub.edu}

\affil[1]{\orgdiv{Small Biosystems Lab, Condensed Matter Physics Departement}, \orgname{Universitat de Barcelona}, \orgaddress{\street{C/ Marti i Franques 1}, \city{Barcelona}, \postcode{08028}, \country{Spain}}}

\affil[2]{\orgdiv{Unit of Biophysics and Bioengineering, Department of Biomedicine, School of Medicine and Health Sciences}, \orgname{Universitat de Barcelona}, \orgaddress{\street{C/Casanoves 143}, \city{Barcelona}, \postcode{08036}, \country{Spain}}}

\affil[3]{\orgdiv{Institute for Bioengineering of Catalonia (IBEC)}, \orgname{The Barcelona Institute for Science and Technology (BIST)}, \orgaddress{\city{Barcelona}, \postcode{08028}, \country{Spain}}}

\affil[4]{\orgname{Steven B. Smith Engineering}, \orgaddress{\city{Los Lunas}, \country{New Mexico, USA}}}

\affil[5]{\orgname{Institut de Nanoci\`encia i Nanotecnologia (IN2UB)}, \orgaddress{\city{Barcelona}, \country{Spain}}}

%----------------------------------------------------------------------------
%****************************************************************************
%----------------------------------------------------------------------------

\abstract{DNA hybridization is a fundamental reaction with wide-ranging applications in biotechnology.  The nearest-neighbor (NN) model provides the most reliable description of the energetics of duplex formation. Most DNA thermodynamics studies have been done in melting experiments in bulk, of limited resolution due to ensemble averaging. In contrast, single-molecule methods have reached the maturity to derive DNA thermodynamics with unprecedented accuracy. We combine single-DNA mechanical unzipping experiments using a temperature jump optical trap with machine learning methods and derive the temperature-dependent DNA energy parameters of the NN model. In particular, we measure the previously unknown ten heat-capacity change parameters $\Delta C_p$, relevant for thermodynamical predictions throughout the DNA stability range. Calorimetric force spectroscopy establishes a groundbreaking methodology to accurately study nucleic acids, from chemically modified DNA to RNA and DNA/RNA hybrid structures.}

\keywords{DNA thermodynamics, DNA heat capacity change, Single-molecule force spectroscopy}

\maketitle

%****************************************************************************

Nucleic acid (NA) thermodynamics is essential to understand duplex formation, where two single strands of DNA or RNA hybridize to form a double helix. Hybridization is a crucial process for genome maintenance with many biotechnological applications, from PCR amplification to gene editing \cite{ivanov2020cas9} and DNA origami \cite{castro2011primer, funke2016uncovering}. Accurate knowledge of the thermodynamic energy parameters of NA hybridization is necessary for developing better protocols, often involving heating and cooling cycles for dissociating and hybridizing complementary strands. Most hybridization studies involve calorimetric melting experiments in bulk, where signals such as heat, UV light absorbance, and fluorescence are measured over samples typically containing $10^{10}$ molecules in aqueous solutions \cite{chalikian1999more, vologodskii2018dna}. 

Despite recent progress, basic questions about NA hybridization remain unanswered, such as the nature of the transition state and duplex stability far from standard conditions. Examples are extreme cold and high temperatures, molecular condensates, and confined spaces. 
Thermodynamic predictions far from standard conditions require far-fetched extrapolations of the currently known energy parameters. 
The heat-capacity change at constant pressure, $\Delta C_p$, is crucial for NA formation. $\Delta C_p$ quantifies the temperature dependence of enthalpy ($\Delta H$) and entropy ($\Delta S$) contributions to the free energy of hybridization through the thermodynamic relations at constant pressure, $\Delta C_p=\partial \Delta H/\partial T=T\partial \Delta S/\partial T$. From a microscopic viewpoint, $\Delta C_p$ relates to the change in the number of degrees of freedom, $\Delta n$, in hybridization reactions according to the equipartition law, $\Delta C_p=k_B\Delta n/2$, with $\Delta n = 1$ per cal mol$^{-1}$K$^{-1}$ unit of $\Delta C_p$. It has been suggested that the most significant contribution to $\Delta C_p$ in duplex formation occurs in the alignment of the complementary single-strands upon hybridization \cite{whitley2017elasticity}. 

The temperature dependence of the enthalpy and entropy of DNA hybridization has been neglected for a long time. The assumption $\Delta C_p=0$ was mainly adopted during the first scanning calorimetry studies that could not detect $\Delta C_p$ \cite{breslauer1986predicting}. Over time, improvements in calorimetric measurements \cite{breslauer2021shaping} pointed out the significant role $\Delta C_p$ played in DNA hybridization. During the last decades, bulk \cite{rouzina1999heatI, rouzina1999heatII, wu2002temperature, mikulecky2006heat1, mikulecky2006heat2, dragan2019thermodynamics, volker2020heat} and single-molecule \cite{williams2001entropy, de2015temperature, nicholson2021measuring} experiments assessed the effects of temperature, yielding $\Delta C_p$ values per bp spanning two orders of magnitude depending on the experimental condition, the technique used and the DNA sequence. Single-molecule methods such as atomic force microscopy \cite{rief1999sequence, krautbauer2003unzipping}, FRET \cite{bacic2020recent}, and optical tweezers \cite{bustamante2003ten, moffitt2008recent} have now reached the maturity to address such challenges. While force spectroscopy derives free energy differences ($\Delta G$) from work measurements, single-molecule FRET does it from the lifetimes of states using the Boltzmann-Gibbs distribution \cite{roy2008practical}. The mechanical unzipping of single NA hairpins permits measuring the folding free energy landscape along the reaction coordinate defined by the number of released nucleotides during the unzipping process  \cite{huguet2009statistical, huguet2010single}. The unzipping reaction finds applications in the footprinting of DNA-binding restriction enzymes \cite{koch2002probing}, transcription factors \cite{rudnizky2018single, khamis2021single}, and peptides \cite{camunas2015single}. Mechanical unzipping has also permitted the design of calipers for measuring molecular distances \cite{shrestha2021single}. Here, we derive the elusive heat capacity change $\Delta C_p$ for the different DNA nearest-neighbor base pair motifs.

DNA's thermodynamic stability ($\Delta G$) results from the compensation of the favorable $\Delta H$ and the unfavorable $\Delta S$ of folding, $\Delta G = \Delta H - T\Delta S$, with $T$ the temperature. For $\Delta C_p=0$, $\Delta H$ and $\Delta S$ are $T$-independent, and $\Delta G$ is linear with $T$. For $\Delta C_p\neq 0$, enthalpy-entropy compensation makes $\Delta H$ and $T\Delta S$ of comparable magnitude masking deviations of $\Delta G$ from a $T$-linear behavior. Potentially, one could derive the temperature-dependent $\Delta S$ from $\Delta G$ using the relation $\Delta S=-\partial \Delta G/\partial T$. However, this method is imprecise due to strong compensation between $\Delta H$ and $T\Delta S$. To determine $\Delta C_p$, it is convenient to measure either enthalpy or entropy contributions independently of $\Delta G$. We introduce a method to accurately derive DNA thermodynamics by directly measuring the temperature-dependent entropy of hybridization $\Delta S$ using calorimetric force spectroscopy with optical tweezers \cite{de2015temperature}. We apply the Clausius-Clapeyron equation to single-molecule experiments \cite{rico2022molten} and combine it with a tailored machine-learning algorithm. This approach allows us to derive the entropies, enthalpies, and $\Delta C_p$ parameters of hybridization at single base pair resolution in the nearest-neighbor (NN) model \cite{devoe1962stability, crothers1964theory}. 

According to the NN model, the duplex's free energy, entropy, and enthalpy equal the sum of all contributions of the adjacent nearest-neighbor base pairs (NNBP) along the sequence. The four distinct canonical Watson-Crick base pairs generate sixteen possible NNBP combinations (e.g., AG/TC, meaning that $5^{\prime}-$AG$-3^{\prime}$ hybridizes with $3^{\prime}-$TC$-5^{\prime}$) with their corresponding energy parameters. The sixteen parameters reduce to ten due to strand complementarity symmetry (e.g., AG/TC equals CT/GA), further reducing to eight from circular symmetry relations \cite{gray1970new, licinio2007irreducible, huguet2017derivation}. The ten NNBP DNA parameters have been derived at $37^{\circ}$C from melting experiments of short DNA duplexes by many laboratories worldwide \cite{gotoh1981stabilities, delcourt1991stacking, doktycz1992studies, santalucia1996improved, sugimoto1996improved} and unified by Santalucia \textit{et al.} \cite{santalucia1998unified} in the so-called \textit{Unified Oligonucleotide} (UO) set. In the last decade, the NNBP free energy parameters have been derived from reversible work measurements in mechanical unzipping experiments of DNA and RNA hairpins at room temperature (298K) \cite{huguet2010single, huguet2017derivation, rissone2022stem}. These energy parameters are used by most secondary structure prediction tools, such as Mfold \cite{zuker2003mfold}, Vienna package \cite{lorenz2011viennarna}, uMelt \cite{dwight2011umelt}, among others. Here, we apply calorimetric force spectroscopy to measure the NNBP energy parameters in the temperature range $7-40^{\circ}$C and derive the $\Delta C_p$ values.  

%------------------------------------------------------------------------------
%******************************************************************************
%------------------------------------------------------------------------------

\section*{Results} 

\begin{figure}[t!]
\centering
\includegraphics[width=\textwidth]{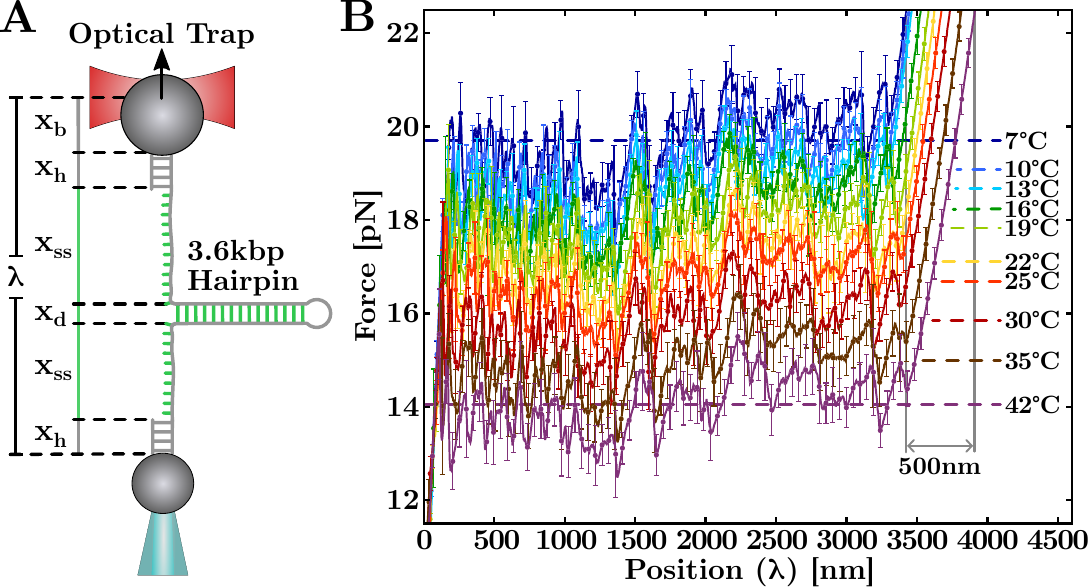}
\caption{\label{fig:FIG1}\textbf{Single-molecule calorimetric force-spectroscopy.} (\textbf{A}) Experimental setup (Sec. \ref{methods:setup}, Methods) showing each component of the (measured) total trap-pipette distance, $\lambda$. (\textbf{B}) Temperature dependence of FDCs obtained by pulling the 3.6kbp DNA hairpin. At each $T$, the FDC results from averaging over $5-6$ molecules for $40-50$ unzipping/rezipping cycles. The error bars (plotted for a fraction of the total data points) show the molecule-to-molecule variability.}
\end{figure}

\begin{figure}[t!]
\centering
\includegraphics[width=\textwidth]{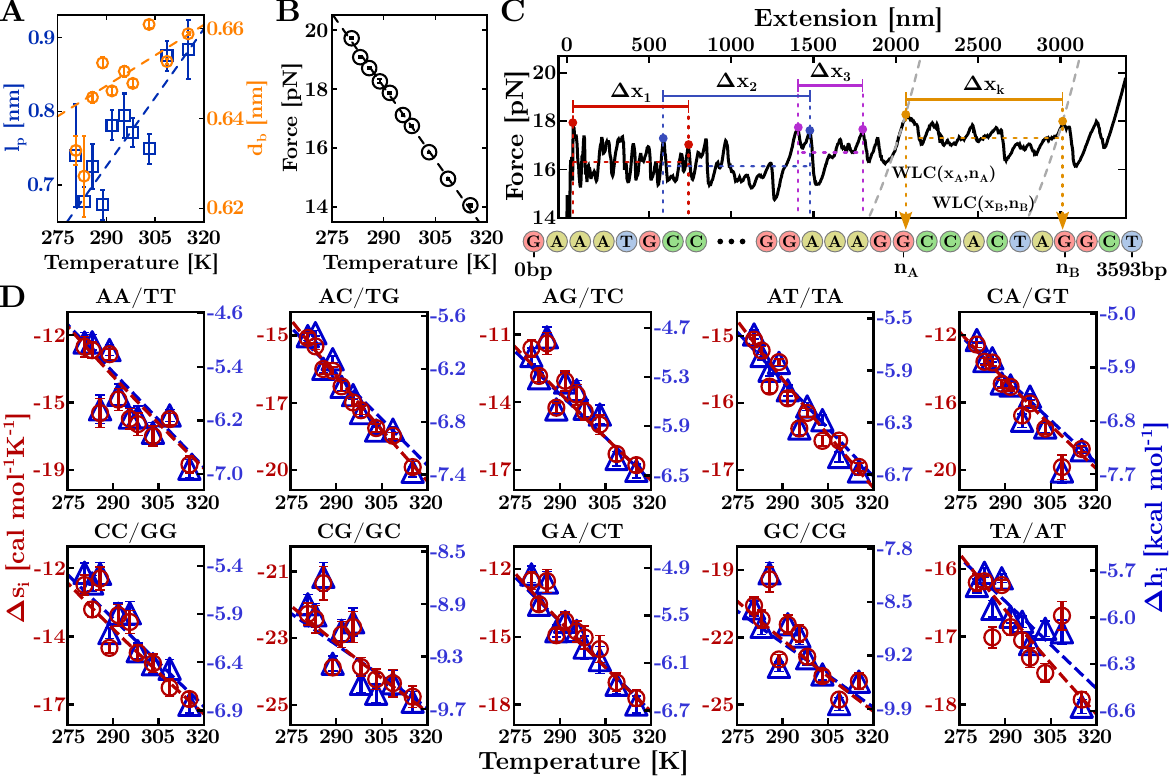}
\caption{\label{fig:FIG2}\textbf{NNBP entropies and enthalpies.} (\textbf{A}, \textbf{B}) $T$-dependence of the persistence length, $l_p$, (blue) interphosphate distance, $d_b$, (orange) and average unzipping force, $f_{\rm m}$ (black). Linear fits to data (black, blue, and orange lines, respectively) are also shown. (\textbf{C}) Example of application of the Clausius-Clapeyron relation to the experimental FDC (see text). (\textbf{D}) $T$-dependence of the ten NNBP entropies (red) and enthalpies (blue). Results are reported in \tcr{Extended Data Tables 2 and 4}, respectively. The entropy of motifs GC/CG and TA/AT were obtained by applying the circular symmetry relations. Fits to data (see text) are shown with a red (entropy) and blue (enthalpy) dashed line.}
\end{figure}

We used a temperature-jump optical trap (Fig. \ref{fig:FIG1}A and Sec. \ref{methods:setup}, Methods) to unzip a 3593bp ($\approx 3.6$kbp) DNA hairpin made ending in a GAAA tetraloop. Pulling experiments were conducted at temperatures $7-42\deg$C at 1M NaCl in 10mM Tris-HCl buffer (pH 7.5). Figure \ref{fig:FIG1}B shows the measured force-distance curves (FDCs). They exhibit a sawtooth pattern upon increasing the trap-pipette distance, $\lambda$, until the hairpin unzips completely and the elastic response of the released single-stranded DNA (ssDNA) is measured (rightmost part of Fig. \ref{fig:FIG2}B). 
Upon increasing $T$, the hairpin unzips at progressively lower forces (horizontal dashed lines in Fig.\ref{fig:FIG1}B) from $\sim 20$pN at $7\deg$C to $\sim 14$pN at $42\deg$C. This indicates that the DNA stability decreases with $T$, and the energy parameters of the NN model are temperature dependent: the higher the temperature, the lower the hairpin's free energy of hybridization. 
Moreover, the molecular extension of the ssDNA at a given force increases with $T$, yielding a total of $\sim + 500$nm between $7\deg$ and $42\deg$C (Fig. \ref{fig:FIG2}B, horizontal grey double arrow).

%------------------------------------------------------------------------

\bigbreak
\noindent
\textbf{$\mathbf{T}$-dependence of the ssDNA elasticity.} Deriving the full NNBP parameters in unzipping experiments requires measuring the $T$-dependent ssDNA elasticity. To do this, we extract the force versus the hairpin's molecular extension ($x$) curve (hereafter referred to as FEC) with 
\begin{equation}
	\label{eq:lambda}
	\lambda=x+x_{b}+2x_h\Rightarrow x=\lambda-x_{b}-2x_h\, ,
\end{equation}
with $x$ being shown as a green vertical line in Fig. \ref{fig:FIG1}A. Here, $x_b$ and $2x_h$ are the bead displacement relative to the trap's center and the handles extension, respectively (grey vertical lines in Fig. \ref{fig:FIG1}A). 
To determine the term $x_b+2x_h$ in Eq.\eqref{eq:lambda}, we have used the \textit{effective stiffness} method \cite{severino2019efficient} with $x_b+2x_h=f/k_{\rm eff}$ and $x=\lambda-f/k_{\rm eff}$ where $k_{\rm eff}$ is obtained from a linear fit to the first slope in the FDC when the hairpin is fully folded (\tcr{Sec. 1, Supp. Info}). Notice that all extensions are $f$ and $T$ dependent.

From Fig. \ref{fig:FIG1}A, $x=x_{\rm ss} + x_d$, where $x_{\rm ss}$ is the ssDNA extension and $x_d$ is the projection of the helix diameter ($d=2$nm \cite{woodside2006direct}) on the pulling axis, which is described by Eq.\eqref{eq:FJC} (Methods). To model the ssDNA elasticity, we use the intextensible WLC model \cite{smith1996overstretching, viader2021cooperativity} (\tcr{Sec. \ref{methods:ssDNAelastic}, Methods}). In this model, the extension $x_{\rm ss}$ at a given force is proportional to the number of released bases $n$, $x_{\rm ss}(f,T)=n x^{(1)}_{\rm ss}(f,T)$ where $x^{(1)}$ is the extension per base. For the fully folded hairpin, $n=0$ and $x=x_d$, whereas for the fully unzipped hairpin, $x= 2x_{\rm ss}=2(N+L/2)x^{(1)}_{\rm ss}$ with $N$ the number of base pairs in the stem and $L$ the loop size. 
At a given $T$, we obtain $x^{(1)}_{\rm ss}(f,T)$ from the FEC measured after the last force rip (\tcr{Extended Data Fig. 2}). A fit of the WLC in Eq.\eqref{eq:WLC} (Methods) to the rightmost part of the FEC gives the temperature-dependent persistence length ($l_p$) and inter-phosphate distance ($d_b$) of the ssDNA (Fig. \ref{fig:FIG2}A and \tcr{Extended Data Table 1}). As $T$ increases, $l_p$ (blue squares), varies from $l_p^{280\rm K} = 0.74(7)$nm to $l_p^{315\rm K} = 0.88(4)$nm ($\approx +30\%$). A linear fit to the data gives the slope $6(1) \cdot 10^{-3}$nm/K (blue line). Moreover, the interphosphate distance, $d_b$, (orange circles) shows a weak linear $T$-dependence ($\approx +5\%$) of slope $4(1) \cdot 10^{-4}$nm/K (orange line). Similar behavior has been observed for shorter ssDNAs of $20-40$ nucleotides \cite{rico2022temperature} and polypeptide chains \cite{rico2022molten}. The observed increase in $x_{\rm ss}$ with $T$ is predicted in Debye-Huckel theory due to the entropy of the cloud of counterions. Upon increasing temperature, the screening of the phosphates repulsion is reduced, and $l_p$ increases.

%----------------------------------------------------------------------------

\bigbreak
\noindent
\textbf{Derivation of the NNBP entropies.} 
To derive the entropies of the different NNBPs, we have decomposed the full unzipping curve into segments of variable length encompassing different regions along the FDC. Each segment is delimited by two peaks corresponding to force rips along the FEC. Figure \ref{fig:FIG2}C shows examples of segments starting and ending at a peak (colored circles). The entropy of hybridization, $\Delta S_{0,k}(T)$, of each segment $k$ is given by (Sec. \ref{methods:CCequation}, Methods),
\begin{equation}
	\label{eq:CC}
	\Delta S_{0,k}(T) =  \frac{\partial f_{\rm m,k}(T)}{\partial T} \Delta x_k(f_{\rm m,k}(T),T) + \int_0^{f_{\rm m,k}(T)} \frac{\partial \Delta x_k(f,T)}{\partial T}df \, ,
\end{equation}
with $\Delta x_k$ the extension of segment $k$. Equation \eqref{eq:CC} is analogous to the Clausius-Clapeyron equation \cite{smith1996overstretching, de2015temperature, rico2022molten} in classical thermodynamics. Here $f$ and $x$ stand for the equivalent quantities of pressure and volume in hydrostatic systems. The r.h.s. of Eq.\eqref{eq:CC} depends on the average unzipping force of segment $k$ measured at different temperatures, $f_{\rm m,k}(T)$, according to the equal area Maxwell construction for segment $k$ (colored horizontal dashed lines in Fig. \ref{fig:FIG2}C). $f_{\rm m,k}(T)$ varies linearly with $T$, all segments showing the same slope $-0.165(3)$pN/K within statistical errors (Fig. \ref{fig:FIG2}B \tcr{and Extended Data Table 1}). The integral in Eq.\eqref{eq:CC} accounts for the work needed to stretch the ssDNA and orient the molecule along the pulling axis between zero force and $f_{\rm m,k}(T)$ (\tcr{Fig. 3A and Table 1, Extended Data}). Equation \eqref{eq:CC} applied to the full FEC gives the total entropy of hybridization of the hairpin (\tcr{Extended Data Fig. 3B}). 

\begin{figure}[t!]
\centering
\includegraphics[width=\textwidth]{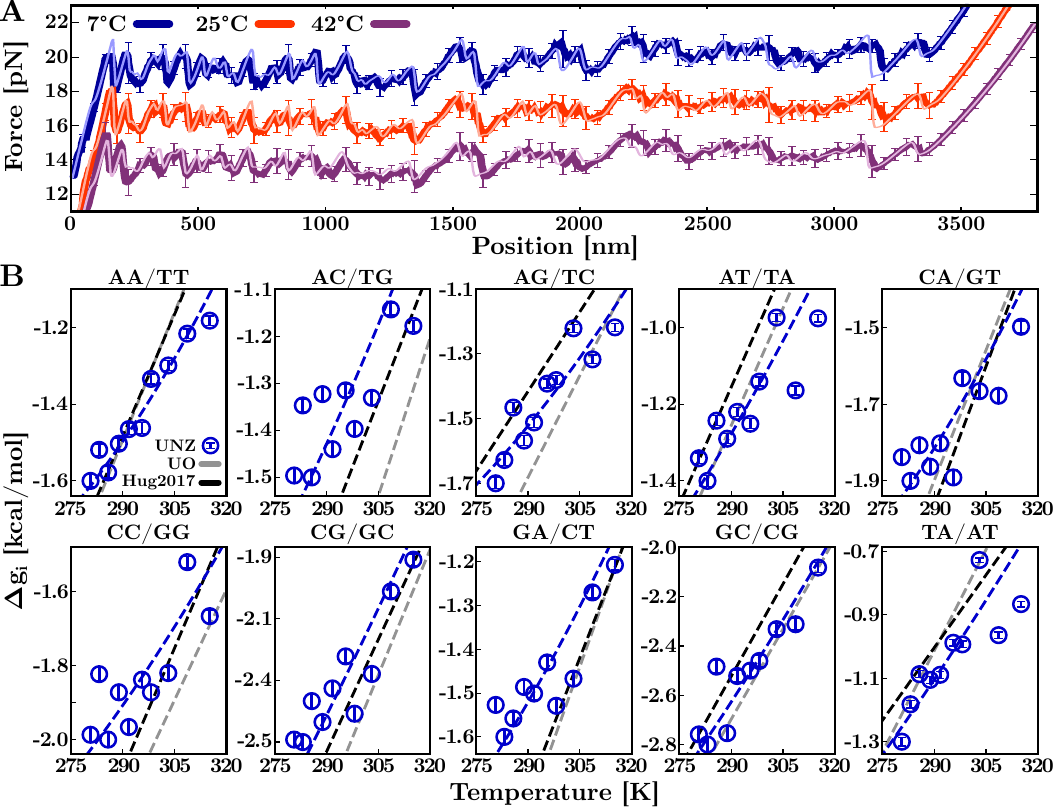}
\caption{\label{fig:FIG3}\textbf{T-dependence of the NNBP DNA free energies.} (\textbf{A}) Experimental FDCs (dark-colored lines) and theoretical predictions (light-colored lines) at $7^{\circ}$C, $25^{\circ}$C, and $42^{\circ}$C. Analogous results have been obtained at all temperatures (\tcr{Extended Data Fig. 5}). (\textbf{B}) Results for the ten NNBP DNA free energies (\tcr{Extended Data Table 3}). The free energy of motifs GC/CG and TA/AT has been computed with circular symmetry relations. A fit to data (blue line) has been added to compare with predictions by the UO (grey line) and Huguet \textit{et al.} \cite{huguet2017derivation} (black line) sets.}
\end{figure}

To apply Eq.\eqref{eq:CC} for a given segment $k$, we must identify the DNA sequence limited by the initial and final peaks. A WLC curve passing through a peak at $(x,f)$ gives the number $n$ of unzipped bases at that peak (dashed-grey lines in segment $\Delta x_k$). The initial and final values $n_A$ and $n_B$ (orange segment in Fig. \ref{fig:FIG2}C), identify the DNA sequence of that segment.
Let $k=1,2,\dots, K$ enumerate the different segments. The entropy of segment $k$ at zero force and temperature $T$ in the NN model is given by the sum of the individual entropies of all adjacent NNBPs within that segment, 
\begin{equation}
	\label{eq:NNBP_CC}
	\Delta S_{0,k}(T) = \sum_{i={\rm AA, CA, \dots}} c_{k,i} \Delta s_{i}(T)\, ,
\end{equation}
where the sum runs over the ten independent NNBP parameters labeled by the index $i$, and $\Delta s_{i}$ is the entropy of motif $i$ with multiplicity $c_{k,i}$, i.e., the number of times motif $i$ appears in segment $k$. The entropy $\Delta S_{0,k}(T)$ in the l.h.s of Eq.\eqref{eq:NNBP_CC} is measured using Eq.\eqref{eq:CC}, and $c_{k,i}$ for each motif is obtained from the segment sequence. 
A stochastic gradient descent algorithm has been designed to solve the system of $K$ non-homogeneous linear equations \eqref{eq:NNBP_CC} and derive the $\Delta s_{i}$ parameters at each $T$ (Sec. \ref{methods:SGDentropies}, Methods). The results for the $T$-dependent DNA NNBP entropies $\Delta s_{i}$ are shown in Fig. \ref{fig:FIG3}D and reported in \tcr{Extended Data Table 2}. Typically, $K\sim 400-600$, making our single-molecule approach equivalent to melting experiments on different oligo sequences.

%------------------------------------------------------------------------------

\bigbreak
\noindent
\textbf{NNBP free energies and enthalpies.} 
From the previously derived NNBP entropies $\Delta s_i(T)$, we can also derive the NNBP enthalpies, $\Delta h_i(T)$, from the relation,
\begin{equation}
	\label{eq:dH}
	\Delta h_i(T) = \Delta g_i(T) + T\Delta s_i(T) \, ,
\end{equation}
with $\Delta g_i(T)$ the free energies of the different motifs. To measure the $\Delta g_i(T)$, we have fitted the FDCs of Fig. \ref{fig:FIG1}B to the unzipping curves predicted by the NN model \cite{huguet2009statistical, huguet2010single, huguet2017derivation}.
The fitting procedure is based on a Monte-Carlo method that optimizes the eight independent energy parameters, $\Delta g_i(T)$, at each $T$ (\tcr{Sec. \ref{methods:MCenergies}, Methods}). The other two energy parameters (GC/CG and TA/AT) are obtained from the circular symmetry relations \cite{gray1970new, licinio2007irreducible, huguet2017derivation}.
Fits are shown in Fig. \ref{fig:FIG3}A for three selected temperatures, and the $\Delta g_i(T)$ are shown in Fig. \ref{fig:FIG3}B (see also \tcr{Extended Data Table 3}). Results (blue circles) agree with the unified oligonucleotide (UO) dataset (black line) and the energy parameters obtained by Huguet \textit{et al.} in Ref.\cite{huguet2017derivation} (grey line). In this reference, unzipping experiments at room temperature (298K) were combined with melting temperature data of oligo hybridization over the vastly available literature. Overall agreement is observed, except for some motifs such as AC/TG and GA/CT where the UO energies are lower. The ten NNBP enthalpies were obtained from Eq.\eqref{eq:dH} at each $T$ and are shown in Fig. \ref{fig:FIG2}D (see also \tcr{Extended Data Table 4}). The agreement between the new $\Delta g_i(T)$ values in Fig. \ref{fig:FIG3}B with previous measurements under the assumption that $\Delta c_p^i=0$ for all motifs \cite{santalucia1998unified,huguet2017derivation}, underlines the strong compensation between the temperature-dependent enthalpies and entropies shown in Fig. \ref{fig:FIG2}D that mask the finite $\Delta c_p^i$'s.

%---------------------------------------------------------------------------

\begin{figure}[t!]
\centering
\includegraphics[width=\textwidth]{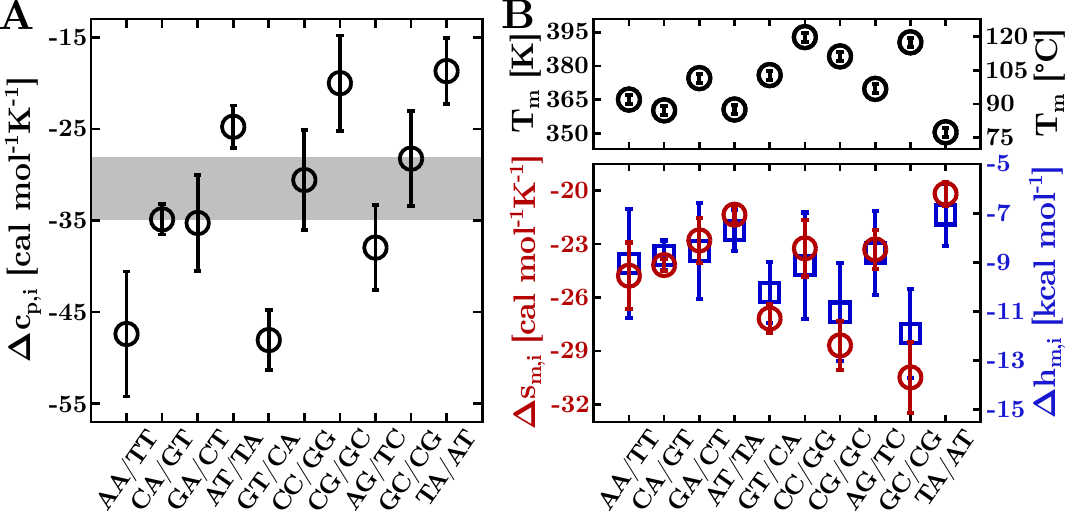}
\caption{\label{fig:FIG4}\textbf{The DNA NNPB thermodynamics.} (\textbf{A}) Measured heat capacity change per motif. The grey band shows the range of $\Delta c_p$ values per motif reported in Ref.\cite{mikulecky2006heat1}. (\textbf{B}) Melting temperatures (top), entropies, and enthalpies at $T_m$ (bottom) for each of the ten NNBP parameters. Results for motifs GC/CG and TA/AT have been derived by applying circular symmetry relations.}
\end{figure}

\bigbreak
\noindent
\textbf{NNBPs heat capacity changes.} 
The temperature-dependent NNBP entropies and enthalpies permit us to derive the heat capacity changes $\Delta c_{p,i}$ for all motifs by using the relations,  
\begin{subequations}
	\label{eq:dSdHfit}
	\begin{align}
		\label{eq:dSfit}
		\Delta s_i &= \Delta s_{m,i} + \Delta c_{p,i} \log(T/T_{m,i}) \\
		\label{eq:dHfit}
		\Delta h_i &= \Delta h_{m,i} + \Delta c_{p,i} (T-T_{m,i}) \, ,
	\end{align}
\end{subequations}
where $T_{m,i}$ is the melting temperature of motif $i$ where $\Delta g_i(T_{m,i})=0$, and $\Delta s_{m,i}$ and $\Delta h_{m,i}$ are the entropy and enthalpy at $T_{m,i}$, fulfilling $\Delta h_{m,i}=T_{m,i}\Delta s_{m,i}$. To derive the ten $\Delta c_{p,i}$, we fit the NNBP entropies to the equation $A_i + \Delta c_{p,i} \log(T)$, being $A_i= \Delta s_{m,i} - \Delta c_{p,i} \log(T_{m,i})$. The results are shown in Fig. \ref{fig:FIG4}A and Table \ref{tab:NNBPthermo} (column 1). From the $\Delta c_{p,i}$, we combined Eqs.\eqref{eq:dSdHfit} with Eq.\eqref{eq:dH} to fit the experimental values of $\Delta g_i(T)$ (blue dashed lines in Fig. \ref{fig:FIG3}B) to obtain $T_{m,i}$. From the $T_{m,i}$, we retrieve $\Delta s_{m,i}$ and $\Delta h_{m,i}$ from Eqs.\eqref{eq:dSfit}, \eqref{eq:dHfit} (red and blue dashed lines in Fig. \ref{fig:FIG2}D). The fitting procedure is described in \tcr{Sec. 4, Supp. Info}. Results for $T_{m,i}$, $\Delta s_{m,i}$ and $\Delta h_{m,i}$ are shown in Fig. \ref{fig:FIG4}B and Table \ref{tab:NNBPthermo}. Notice the high $T_m$ values of the individual motifs, a consequence of the high enthalpies of the NN motifs.  

%-------------------------------------------------------------------------------
%******************************************************************************
%------------------------------------------------------------------------------

\bigbreak
\noindent
\textbf{Discussion.}
We measured the free energies, entropies, and enthalpies in the temperature range $7-42^{\circ}$C at the level of single nearest-neighbor base pairs (NNBP). We have mechanically unzipped a 3.6kbp DNA hairpin using a temperature-jump optical trap. The DNA sequence is long enough to permit us to accurately derive the ten NNBP free-energy parameters, $\Delta g_{i}$, by statistical modeling of the force-distance curve (FDC) \cite{huguet2010single, huguet2017derivation}. At first sight, the $\Delta g_{i}$ values (Fig. \ref{fig:FIG3}B) vary linearly with temperature due to the compensation of enthalpy and entropy in Eq.\eqref{eq:dH}. This compensation masks the temperature dependence of enthalpies, $\Delta h_i$, and entropies, $\Delta s_i$, (cf. Eqs.\eqref{eq:dSdHfit}) arising from a finite $\Delta C_p$, rendering $\Delta g_{i}=\Delta h_i-T\Delta s_i$ linear in $T$. We have introduced an approach to derive the $T$-dependent entropies by combining the Clausius-Clapeyron relation in force (Eq.\eqref{eq:CC}) and the nearest-neighbor (NN) model for duplex formation. We implemented a tailored stochastic gradient descent algorithm to extract the ten $T$-dependent NNBP entropy parameters, $\Delta s_{i}$. Together with the $\Delta g_{i}$ values, the ten enthalpy parameters, $\Delta h_{i}$, readily follow. Fitting the results to Eqs.\eqref{eq:dSdHfit}, we have obtained the $\Delta c_{p,i}$ and $T_{m,i}$ values for the ten motifs (Fig. \ref{fig:FIG4} and Table \ref{tab:NNBPthermo}). Upon averaging over all motifs we get $\overline{\Delta c_p}= -35 (9)$ cal mol$^{-1}$K$^{-1}$bp$^{-1}$. This must be compared to the highly dispersed results from bulk experiments ranging between -20 and -160 cal mol$^{-1}$K$^{-1}$bp$^{-1}$, depending on the experimental technique, setup, and DNA sequence \cite{rouzina1999heatI, mikulecky2006heat1}. In contrast, recent molecular dynamic simulations estimated an average $\Delta c_p\sim -30$ cal mol$^{-1}$K$^{-1}$bp$^{-1}$ \cite{lomzov2015evaluation, hadvzi2021origin}, in agreement with our results.

Force spectroscopy emerges as a reliable approach to accurately derive the energy parameters in NAs. Unzipping experiments control the unfolding reaction by moving the force-sensing device (e.g., optical trap in optical tweezers and cantilever in AFM). In DNA hairpins of a few kb, the sequence contains all ten NN motifs repeated several times, ensuring their reliable statistical sampling in single-DNA unzipping experiments. The high-temporal resolution combined with the sub-kcal/mol accuracy of work measurements permits us to derive the ten NN energy parameters at different temperatures. The main requirement of unzipping experiments is an accurate model of the elastic response of the single-stranded DNA (ssDNA). We have fitted the last part of the unzipping FDCs to the worm-like chain model, known to fit data well at high-salt conditions (1M NaCl) where ssDNA excluded-volume effects are negligible \cite{viader2021cooperativity}. Salt concentration might also affect the $\Delta c_{p,i}$ values. As these are related to the change in configurational entropy upon duplex formation, salt-dependent $\Delta c_p$'s might indicate a change in conformational heterogeneity in either the dissociated or hybridized strands upon varying salt.

How do our results compare to those derived from calorimetric melting experiments? The structure of the unfolded state differs in unzipping and thermal denaturation experiments, a fully stretched ssDNA at a given force, and a random coil at zero force, respectively. Their free energy difference equals the work to stretch the ssDNA from the random coil to the stretched conformation. Moreover, hybridization and unzipping differ in the order of the unfolding reactions: while hybridization of two complementary oligos, $A$ and $\overline{A}$, is a bimolecular reaction $A+\overline{A} \leftrightharpoons A\overline{A}$, hairpin unzipping is a unimolecular reaction $A \leftrightharpoons B$ between the folded and unfolded conformations. Such difference is apparent in the dependence of $T_m$ on the enthalpy $\Delta H_0^m$ and entropy $\Delta S_0^m$ of folding. For a bimolecular reaction, the value of $T_m$ explicitly depends on the total oligo concentration $c$, with $\Delta S_0^m,\Delta H_0^m$ taken at the reference 1M salt condition (Eq.\eqref{eq:TmBI}, Methods). Instead, for the unimolecular unzipping reaction, $T_m=\Delta H_0^m/\Delta S_0^m$ does not include the entropy of mixing the dissociated strands. We expect that temperature-dependent enthalpies $\Delta h_i$ are equal for hybridization and unzipping, whereas entropies $\Delta s_i$ should differ due to the entropy of mixing. We have determined the homogeneous entropy correction $\delta\Delta s$ to the total entropy $\Delta S_0^m$ between hybridization (bimolecular) and unzipping (unimolecular) reactions (\tcr{Sec. \ref{methods:Tmelting}, Methods}). The correction is an intensive quantity that is independent of oligo sequence and length, $\delta\Delta s= 6(1)\, {\rm cal \, mol^{-1}K^{-1}} \sim 4R\log 2$, where $R=1.987 \, {\rm cal \, mol^{-1}K^{-1}}$ is the ideal gas constant (Eq.\eqref{eq:TmUNI}, Methods). This value has been obtained by comparing the $T_m$ values predicted by our energy parameters using Eq.\eqref{eq:TmBI} with the experimental values obtained for DNA duplexes in Ref.\cite{owczarzy2008predicting} at 1020mM NaCl and $c=2\mu$M. The results of such a comparison are shown in \tcr{Extended Data Fig. 6}. Practically, the effect of the entropic correction $\delta\Delta s$ on $T_m$ is small as it is $\sim 3-4$ times lower than the average NNBP's entropy $\overline{\Delta s_{m}}\sim -20$ cal mol$^{-1}$K$^{-1}$ (cf. Table \ref{tab:NNBPthermo}). Notice that $\Delta S_0^m$ is extensive, growing linearly with the oligo length, whereas $\delta\Delta s$ is intensive. Therefore, the correction $\delta\Delta s\sim 5$ cal mol$^{-1}$K$^{-1}$ is negligible for sufficiently large oligos, being already $5\%$ for oligos of just ten nucleotides and further decreasing for longer DNAs (\tcr{see Extended Data Table 5}).

%-----------------------------------------------------------------------------
%*****************************************************************************
%-----------------------------------------------------------------------------

\bigbreak
\noindent
\textbf{Conclusions.}
The remarkable accuracy of the nearest-neighbor model for reproducing the experimentally measured force-distance curves permitted us to derive the temperature-dependent DNA energy parameters. One might ask whether there are deviations from the NN model, e.g., in the form of next-to-nearest neighbor (NNN) effects predicted to be important for some tetranucleotide motifs \cite{balaceanu2019modulation}. However, NNN effects might be difficult to observe in unzipping experiments of long DNA hairpins. Our method averages local effects over the whole sequence hiding potential deviations from the NN model at some locations. Unzipping studies on suitably designed short DNA hairpins containing specific NNN motifs would be more appropriate to address this problem. In this case, determining the elastic response of the specific ssDNA sequence would also be necessary \cite{alemany2014determination}. The unzipping method might also be applicable to derive the temperature-dependent energy parameters of RNA, where finite $\Delta C_p$ effects are particularly relevant to the RNA folding problem. Previous studies at room temperature show that RNA unzipping is an irreversible process driven by stem-loops forming along the unpaired strands that compete with the hybridization of the native stem \cite{rissone2022stem}. Such irreversible and kinetic effects suggest higher $\Delta c_{p,i}$ values in RNA compared to DNA.
DNA thermodynamics down to $0\deg$C might find applications for predicting DNA thermodynamics at low temperatures and cold denaturation effects. Estimates based on our $\Delta c_{p,i}$ values show that the most stable DNA hairpins ending a tetraloop predict cold denaturation temperatures lower than $\approx -90^{\circ}$C (\tcr{Extended Data Fig. 7}), raising questions about the importance of cold denaturation effects for cryophile organisms surviving in extremely cold environments \cite{d2006psychrophilic}. 
Finally, our results have implications for determining the DNA force fields used in molecular dynamics simulations of DNA conformational kinetics \cite{liao2019long}, essential for computational studies of NAs in general.   

%-----------------------------------------------------------------------------
%*****************************************************************************
%-----------------------------------------------------------------------------

\section*{Methods}

\subsection{Molecular Construct and Experimental Setup}
\label{methods:setup}

We used a temperature-jump optical trap \cite{de2015temperature} (OT) to unzip a 3593bp DNA hairpin flanked by short (29bp) DNA handles and ending with a GAAA tetraloop. Experiments have been carried out in the temperature range [7,42]$^{\deg}$C in a buffer of 1M NaCl, 10mM Tris-HCl (pH 7.5), and 1mM EDTA. Experiments have been performed at a constant pulling speed, $v=100$nm/s. We sampled 5-6 different molecules at each temperature, collecting a minimum of $\sim 50$ unfolding-folding trajectories per molecule.
To change the temperature inside the microfluidics chamber, the MiniTweezers setup implements a heating laser of wavelength $\lambda = 1435$nm. This device allows for increasing the temperature by discrete amounts of $\Delta T \sim +2.5^{\circ}$C up to a maximum of $\sim +30^{\circ}$C with respect to the environment temperature, $T_0$. By placing the OT in an icebox cooled down at a constant $T_0 \sim 5^{\circ}$C, it has been possible to carry out experiments at a minimum temperature of $7^{\circ}$C. The design of the microfluidics chamber has been chosen to damp convection effects caused by the laser non-uniform temperature, which may produce a hydrodynamics flow between medium regions (water) at different $T$. 

In a typical OT unzipping experiment, the molecule is tethered between two polystyrene beads through specific interactions with the molecular ends. One end is labeled with a digoxigenin (DIG) tail and binds with an anti-DIG coated polystyrene bead (AD) of diameter $3\mu$m. The other molecular end is labeled with biotin (BIO) and binds with a streptavidin-coated bead (SA) of diameter $2\mu$m. The SA bead is immobilized by air suction at the tip of a glass micropipette, while the AD bead is optically trapped.
A pulling cycle consists of moving the optical trap between two fixed positions: the molecule starts in the folded state, and the trap-pipette distance ($\lambda$) is increased, resulting in an external force to be applied to the molecular ends. This causes the hairpin to progressively unzip in a stick-slip process characterized by a sequence of slopes and force rips (FDC sawtooth pattern). When the molecule is completely unzipped, the rezipping protocol starts; $\lambda$ is decreased, and the molecule folds back reversibly to the hairpin (native) state. 

%--------------------------------------------------------------------------

\subsection{ssDNA elastic model}
\label{methods:ssDNAelastic}

The total hairpin extension with $n$ unzipped bases at force $f$ and temperature $T$, $x(n,f,T)$, is given by the sum of the ssDNA extension ($x_{\rm ss}(f,n,T)$) plus the contribution of the double helix diameter ($x_d(f, T)$). The ssDNA elastic response has been modeled according to the Worm-Like chain (WLC), which reads
\begin{equation}
    \label{eq:WLC}
    f(x,n,T) = \frac{k_{\rm B}T}{4 l_p} \left[\left(1-\frac{x}{nd_b}\right)^{-2} - 1 + 4\frac{x}{nd_b} \right] \, ,
\end{equation}
where $k_{\rm B}$ is the Boltzmann constant, and $T$ is the temperature, $l_p$ is the persistence length, $d_b$ is the interphosphate distance and $n$ is the number of ssDNA bases. Notice that the computation of the ssDNA extension requires inverting Eq.\eqref{eq:WLC} \cite{severino2019efficient}.

The observed increase in the ssDNA extension with $T$ at a given force (Fig. \ref{fig:FIG1}B) demonstrates that $l_p$ and $d_b$ are $T$-dependent. This contrasts with the original assumption in the WLC model that $l_p$ and $d_b$ are temperature independent and $x_{\rm ss}$ decreasing with $T$ at a given force. Remarkably, Eq.\eqref{eq:WLC} accurately describes our data as stacking effects in mixed purine-pyrimidine sequences are negligible in our experimental conditions\cite{viader2021cooperativity}. 

The contribution of the hairpin diameter to $x$ is given by the projection of the helix diameter in the direction of propagation of the force. It is described as a free dipole in an external force field and is modeled by the FJC model, which reads
\begin{equation}
    \label{eq:FJC}
    x_d(f,T) = d \left[ \coth \left( \frac{fd}{\kt}\right) - \frac{\kt}{fd} \right] \, ,
\end{equation}
where $d=2$nm is the hairpin diameter \cite{woodside2006direct}. 

%-----------------------------------------------------------------------

\subsection{The Clausius-Clapeyron Equation}
\label{methods:CCequation}

In the unzipping experiment, the total trap-pipette distance is given by $\lambda = x + x_b + 2x_{h}$, where $x$ is the extension of the ssDNA plus the molecular diameter, while $x_b$ and $2x_{h}$ are the bead displacement relative to the trap’s center and the handles extension, respectively. 
Starting with the hairpin totally folded, unzipping consists in converting the double-stranded DNA into ssDNA until it is completely unfolded. The free energy needed to fold back the ssDNA into the hairpin and decrease the applied force to zero is given by
\begin{equation}
	\label{eq:dG0}
	\Delta G_0(T) = -\int_0^{f_{\rm m}(T)}  \Delta \lambda(f,T) df \, ,
\end{equation}
where $\Delta \lambda$ is the total extension change of the hairpin between the initial and final states of the unzipping integrated between zero and the mean unzipping force, $f_m$. Notice that the contributions from $x_b$ and $2x_{h}$ to $\lambda$ remain constant during the entire unzipping process so that $\Delta \lambda = \Delta x$, i.e., equals the extension change due to the ssDNA and the molecular diameter only. 

The folding entropy can be directly derived from the thermodynamic relation $\Delta S_0 =-\partial \Delta G_0/\partial T$, which gives
\begin{equation}
	\label{eq:CC2}
	\Delta S_0(T) =  \frac{\partial f_{\rm m}(T)}{\partial T} \Delta x(f_{\rm m}(T),T) + \int_0^{f_{\rm m}(T)} \frac{\partial \Delta x(f,T)}{\partial T}df \, .
\end{equation}
The first term of Eq.\eqref{eq:CC2} is analogous to the Clausius-Clapeyron equation for first-order phase transitions, where $f$ and $x$ are equivalent to pressure and volume. The integral term in Eq.\eqref{eq:CC2} accounts for the (positive) entropic contribution to stretch the ssDNA and orient the molecular diameter along the pulling axis from zero to force $f_m$.

%-----------------------------------------------------------------------

\subsection{Stochastic Gradient Descent Method}
\label{methods:SGDentropies}

The derivation of the DNA NNBP entropies corresponds to solving the non-homogeneous linear system of $K$ equations and $I=8$ parameters given by Eq.\eqref{eq:NNBP_CC}. To do this, we used a custom-designed stochastic gradient descent (SGD) algorithm. The model is described in \tcr{Sec. 2, Supp. Info}. The application of an optimization algorithm to this problem is made possible by the large data set built accounting for all possible peaks combination, which gives $K \sim 400-600 \gg I$ for each experimental $T$.

%-----------------------------------------------------------------------

\subsection{Derivation of the NNBP Free-Energies}
\label{methods:MCenergies}

Starting with an initial guess of the ten independent $\Delta g_{i}$, a random increment of the energies is proposed at each optimization step, and a prediction of the FDC is generated. The latter is given by the competition of two energy contributions at each position of the optical trap ($x_{\rm tot}$): the energy of the stretched molecular construct acting to unfold the molecule ($\Delta G_{\rm el}(x_{\rm tot})$), and the energy of the hybridized bps keeping the hairpin folded ($\Delta G_0(x_{\rm tot},n)$). At a given $x_{\rm tot}$ and $n_1$ hybridized bp (Fig. \ref{fig:FIG2}A), the hairpin unfolds when $\Delta G_{\rm el}(x_{\rm tot})>\Delta G_0(x_{\rm tot},n_1)$ (force rip) and $\Delta n=n_2-n_1$ bp are released lowering the stretching contribution so that $\Delta G_{\rm el}(x_{\rm tot})<\Delta G_0(x_{\rm tot},n_2)$ (\tcr{see Sec. 3, Supp. Info}). The error in approximating the experimental FDC with the theoretical one, $E = (f_{\rm exp}-f_{\rm theo})^2$, drives a Metropolis algorithm: a change of the energy parameters is accepted if the error difference to the previous step is negative ($\Delta E <0$). Otherwise ($\Delta E>0$), the proposal is accepted if $\exp(-\Delta E/T) < r$ with $r$ a random number uniformly distributed $r \in U(0,1)$. The algorithm continues until convergence (until $\Delta E$ is smaller than a given threshold) or until the maximum number of iterations is reached. The parameters corresponding to the smallest value of $E$ are the optimal NNBP free energies. 

%-----------------------------------------------------------------------

\subsection{Prediction of the Oligos Melting Temperature}
\label{methods:Tmelting}

The melting temperature for non-self-complementary sequences in bimolecular reactions (hybridization) is given by
\begin{equation}
    \label{eq:TmBI}
    T^{Bi}_m = \frac{\Delta H_0^m}{\Delta S_0^m + R\log(c/4)} \, ,
\end{equation}
where $\Delta H_0^m = \sum^N_i \Delta h_i$ ($\Delta S_0^m = \sum^N_i \Delta s_i$) is the total duplex enthalpy (entropy) at $T=T_m$, $R=0.001987 \rm \, kcal \, mol^{-1}K^{-1}$ is the ideal gas constant, $c$ is the experimental oligo concentration.
In contrast, for unimolecular reactions (folding), we found
\begin{equation}
    \label{eq:TmUNI}
    T^{Uni}_m = \frac{\Delta H_0^m}{\Delta S_0^m + R\log(c/4) + \delta \Delta s} \, ,
\end{equation}
where $\delta \Delta s = 4R\log2$ is a correction to the total entropy $\Delta S_0^m$. 
By subtracting the inverse of Eq.\eqref{eq:TmBI} and Eq.\eqref{eq:TmUNI}, one gets
\begin{equation}
    \label{eq:TmDifference}
    \delta \Delta s = \Delta H_0^m \left(\frac{1}{T^{Bi}_m} - \frac{1}{T^{Uni}_m} \right) \, ,
\end{equation}
where the term in parenthesis is computed by subtracting the $T_m$ values predicted by our energy parameters using Eq.\eqref{eq:TmBI} to the experimental dataset measured by Ocwzarzy \textit{et al.} at 1020mM NaCl and $c=2\mu$M \cite{owczarzy2008predicting} (\tcr{Extended Data Fig. 6 and Table 5}). Finally, $\Delta H_0^m$ is determined by using our NNBP enthalpy parameters.

%-----------------------------------------------------------------------

\backmatter

\bmhead{Supplementary information} 
This article has accompanying Supplementary Information and Extended Data. 

\bmhead{Acknowledgments}
P.R. was supported by the Angelo Della Riccia Foundation. M.R.-P. was supported by the Spanish Research Council JDC2022-049996-I. F.R. was supported by Spanish Research Council Grant PID2019-111148GB-I00 and the Institució Catalana de Recerca i Estudis Avançats (F. R., Academia Prize 2018).

\bmhead{Authors' contributions}
P.R. and M.R.-P. carried out the experiments. S.B.S. designed and built the instrument. P.R. and M.R.-P. analyzed the data. P.R., M.R.-P., and F.R. designed the research. P.R. and F.R. wrote the paper.

\bmhead{Competing interests}
Steven B. Smith makes and sells optical tweezers. All other co‐authors have no competing interests.

%%=============================================================================
%%============================= BIBLIOGRAPHY ==================================

\bibliography{Bibliography}% common bib file

%******************************************************************************
%--------------------------- FIGURES & TABLES ---------------------------------
%******************************************************************************

%
\newpage
\clearpage
\newpage
\clearpage
\newpage
\clearpage
\newpage
\clearpage
\newpage
\begin{table}[t!]
\centering
%\begin{adjustbox}{width=0.85\textwidth}
    \begin{tabular}{c|cccc}
    \multirow{2}{*}{\textbf{NNBP}} & $\mathbf{\Delta c_{p,i}}$ & $\mathbf{\Delta s_{m,i}}$ & $\mathbf{\Delta h_{m,i}}$ & \multirow{2}{*}{$\mathbf{T_{m,i}}$ \textbf{[K]}} \\
     & \textbf{[cal mol$^{-1}$K$^{-1}$]} & \textbf{[cal mol$^{-1}$K$^{-1}$]} & \textbf{[kcal mol$^{-1}$]}\\
    \toprule
    \textbf{AA/TT} & $-47 (7)$ & $-24.8 (1.8)$ & $-9 (2)$ & $365$\\
    \textbf{AC/TG} & $-35 (2)$ & $-24.2 (0.3)$ & $-8.7 (0.6)$ & $360$\\
    \textbf{AG/TC} & $-35 (5)$ & $-22.8 (1.3)$ & $-8.5 (1.0)$ & $375$\\
    \textbf{AT/TA} & $-25 (2)$ & $-21.3 (0.5)$ & $-7.7 (0.8)$ & $361$\\
    \textbf{CA/GT} & $-48 (3)$ & $-27.2 (0.8)$ & $-10.2 (1.2)$ & $376$\\
    \textbf{CC/GG} & $-31 (5)$ & $-23.2 (1.6)$ & $-9 (2)$ & $393$\\
    \textbf{CG/GC} & $-20 (5)$ & $-28.7 (1.4)$ & $-11 (2)$ & $384$\\
    \textbf{GA/CT} & $-38 (4)$ & $-23.3 (1.1)$ & $-8.6 (1.7)$ & $370$\\
    \midrule
    \textbf{GC/CG} & $-28 (5)$ & $-30 (2)$ & $-12 (2)$ & $390$\\
    \textbf{TA/AT} & $-19 (3)$ & $-20.2 (0.7)$ & $-7.0 (1.3)$ & $350$\\
    \bottomrule
    \end{tabular}
%\end{adjustbox}
\caption{\label{tab:NNBPthermo}Measured heat capacity change ($\Delta c_{p,i}$), entropy ($\Delta s_{m,i}$), enthalpy ($\Delta h_{m,i}$), and melting temperature ($T_{m,i}$) for the ten NNBP motifs. Errors are reported in brackets. Motifs GC/CG and TA/AT have been derived by applying the circular symmetry relations.}
\end{table}
\clearpage

\end{document}

\title[Supplementary Information: DNA Calorimetric Force Spectroscopy at Single Base Pair Resolution]{Supplementary Information: DNA Calorimetric Force Spectroscopy at Single Base Pair Resolution}

\author[1]{\fnm{P.} \sur{Rissone}}\email{rissone.paolo@gmail.com}

\author[2,3]{\fnm{M.} \sur{Rico-Pasto}}\email{m.rico.pasto@gmail.com}

\author[4]{\fnm{S. B.} \sur{Smith}}\email{stevesmith@gmail.com}

\author*[1,5]{\fnm{F.} \sur{Ritort}}\email{ritort@ub.edu}

\affil[1]{\orgdiv{Small Biosystems Lab, Condensed Matter Physics Departement}, \orgname{Universitat de Barcelona}, \orgaddress{\street{C/ Marti i Franques 1}, \city{Barcelona}, \postcode{08028}, \country{Spain}}}

\affil[2]{\orgdiv{Unit of Biophysics and Bioengineering, Department of Biomedicine, School of Medicine and Health Sciences}, \orgname{Universitat de Barcelona}, \orgaddress{\street{C/Casanoves 143}, \city{Barcelona}, \postcode{08036}, \country{Spain}}}

\affil[3]{\orgdiv{Institute for Bioengineering of Catalonia (IBEC)}, \orgname{The Barcelona Institute for Science and Technology (BIST)}, \orgaddress{\city{Barcelona}, \postcode{08028}, \country{Spain}}}

\affil[4]{\orgname{Steven B. Smith Engineering}, \orgaddress{\city{Los Lunas}, \country{New Mexico, USA}}}

\affil[5]{\orgname{Institut de Nanoci\`encia i Nanotecnologia (IN2UB)}, \orgaddress{\city{Barcelona}, \country{Spain}}}

\maketitle

%Rename page headers
\pagestyle{myheadings}
\markboth{SUPPLEMENTARY INFORMATION}{SUPPLEMENTARY INFORMATION}

\tableofcontents

% =====================================================================
% *********************************************************************
% =====================================================================

\renewcommand{\thesection}{Supplementary Methods}% Remove section references...
\section{}

% =====================================================================

\subsection{Temperature Dependence of the DNA FDCs}
\label{sec:DNAelastic}

The elastic properties of ssDNA are strongly temperature dependent (see \tcr{Fig. 1B, main text}). Accurately measuring these properties requires modeling all contributions to the trap-pipette distance, $\lambda$, which includes the optical trap displacement ($x_{\rm b}$), the dsDNA handles ($x_{\rm h}$), the ssDNA ($x_{\rm ss}$), and the molecular diameter ($d_0$).
The setup contributions ($x_{\rm b}$ and $x_{\rm h}$) can be evaluated by using the \textit{effective stiffness} method\tcb{$\,^{48}$}. According to it, these terms are approximated by an effective stiffness, $k_{\rm eff}^{-1} \approx k_{\rm h}^{-1} + k_{\rm b}^{-1}$.
The use of short handles ($29$bp) makes the evaluation of the stretching terms easier as their stiffness is much larger as compared to the trap stiffness ($k_{\rm h} \gg k_{\rm b}$), implying that $k_{\rm eff} \approx k_{\rm b}$. Moreover, if the force varies in a relatively narrow range ($f_{\rm max} - f_{\rm min} \lesssim 10$pN), the trap stiffness can be considered nearly force-independent so $k_{\rm eff}$ is constant along the folded branch of the FDC. 
Therefore, we can estimate $k_{\rm eff}$ by fitting the slope preceding the first force rip in the FDC to the linear equation $f = k_{\rm eff}x$ (orange dashed-line in \tcr{Extended Data Fig. 1}). This allows us to compute the (effective) contribution of the handles and optical trap, $x_{\rm eff}$, to the total trap-pipette distance, $\lambda$.

%NOTE: Ref 48 is severino2019efficient

% =====================================================================

\subsection{Stochastic Gradient Descent in a Nutshell}
\label{subsec:SGD}

The basic principle behind stochastic approximation can be backtracked to the Robbins--Monro algorithm of the $1950s$ \cite{robbins1951stochastic}. Since then, stochastic gradient descent (SGD) methods have become one of the most widely used optimization methods  \cite{curry1944method, polyak1987introduction, qian1999momentum, bottou2010large, bottou2012stochastic}.
SGD is an iterative method for optimizing an objective function, $J(w)$, with suitable smoothness properties (e.g., differentiable or subdifferentiable). The set of parameters, $w^*$, minimizing $J(w)$, is iteratively approximated according to an update algorithm proportional to the antigradient of the objective function, $-\nabla_w J(w)$. Starting from an initial guess of $w$, at each step of the algorithm, the parameters are updated according to
\begin{equation}
	\label{eq:SGDstep}
	\begin{cases}
	w_{t+1} &= w_{t} + v_{t+1} \\
	v_{t+1} &= \beta v_{t} - \eta \nabla_{w_t} J(w)  \, ,
	\end{cases}
\end{equation}
where $v_t$ is the \textit{velocity} of the optimization and $\eta \geq 0$ is the step size (called \textit{learing rate}). The parameter $\beta$ (the so-called \textit{momentum coefficient}) accounts for a fraction of the previous step in the current update.
The critical difference between SGD and standard gradient descent algorithms is that information (total entropy and coefficients) from only one FEC segment ($\Delta x_k$) at a time is used to calculate the step, and the segment is picked randomly at each step.

The SGD convergence rate can be improved by considering Nesterov’s Accelerated Gradient (NAG), introduced in 1983 \cite{nesterov1983method, achlioptas2019stochastic}. According to NAG, the update equations are
\begin{equation}
	\label{eq:NAGstep}
	\begin{cases}
	w_{t+1} &= w_{t} + v_{t+1} \\
	v_{t+1} &= \beta v_{t} - \eta \nabla_{w_t + \beta v_t} J(w)  \, .
	\end{cases}
\end{equation}
While the classic momentum (CM) algorithm updates the velocity vector by computing the gradient at $w_t$, the NAG algorithm computes the gradient at $w_t + \beta v_t$. To make an analogy, while CM faithfully trusts the gradient at the current iteration, NAG puts less faith in it and looks ahead in the direction suggested by the velocity vector; it then moves in the direction of the gradient at the look-ahead point. If $\nabla_{w_t + \beta v_t} J(w) \approx \nabla_{w_t} J(w) $, then the two updates are similar.
The advantage of using NAG is that it converges at a rate of $\mathcal{O}(1/t^2)$, while CM converges at a rate of $\mathcal{O}(1/t)$.

To derive the DNA NNBP entropies from unzipping experiments, we used an SGD minimization implementing NAG update equations. Let us rewrite \tcr{Eq.(2)} (main text) as $\mathbf{\Delta S_0}  = C \mathbf{\Delta s}$, where $\mathbf{\Delta S_0}$ is the vector of entropies measured with the Clausius-Clapeyron equation for each of the $K$ FEC segments, $\mathbf{\Delta s}$ is the vector of the $I=8$ NNBP entropy parameters, and $C$ is the $K \times I$ matrix of the coefficients, $c_{k,i}$.

Thus, for a given loss function (ex., least squares), the algorithm has to minimize
\begin{equation}
	\label{eq:SGD}
	J(w) = \sum_{k=1}^K (\hat w_k - w_k)^2 = \sum_{k=1}^K (\Delta S_{0,k} - C_k \mathbf{\Delta s})^2 \, .
\end{equation}
By using this method, we measured the DNA entropies at the single base-pair level for each experimental temperature in the range $[280,315]$ K (see results in \tcr{Fig. 3C}, main text and \tcr{Extended Data Table 3}). 

%------------------------------------------------------------------------
%************************************************************************
%------------------------------------------------------------------------

\subsection{Prediction of the DNA Unzipping Curve}
\label{sec:UNZtheo}

In unzipping experiments, the total trap--pipette distance, $\lambda$, can be written as
\begin{equation}
    \label{eq:x_tot}
    \lambda(f) = x_{\text{b}}(f) + x_{\text{h}}(f) +x_{\text{ss}}(f,n) + x_{\text{d}}(f) \, ,
\end{equation}
where $x_{\text{b}}(f)$ is the displacement of the bead from the center of the optical trap, $x_{\text{h}}(f)$ and $x_{\text{ss}}(f,n)$ account for the extension of the two double-stranded handles and the ssDNA extension, respectively (described with the WLC model, \tcr{Eq.(5), Methods}), and $x_{\text{d}}(f)$ is the projection of the folded hairpin of diameter $d$ (typically $d = 2$nm for DNA and RNA hairpins\tcb{$\,^{49}$}) along the pulling axis \cite{forns2011improving}. It is modeled with the freely-jointed chain in \tcr{Eq.(6), Methods}.
For a given $\lambda$, the total system free energy is given by
\begin{equation}
\begin{split}
    \label{eq:G_eq}
    \Delta G_{\rm tot}(\lambda,n) =& \Delta G_0(n) + \Delta G_\text{b}(x_\text{b}) + \Delta G_\text{h}(x_\text{h}) + \\ &+\Delta G_\text{ss}(x_\text{ss},n) + \Delta G_\text{d}(x_\text{d}) \, ,
\end{split}
\end{equation}
where $\Delta G_0(n) = \sum_i^n \Delta g_{0,i}$, is the hairpin free-energy of hybridization according to the NN model and the other terms are the energy contributions of the corresponding elastic terms in Eq.\eqref{eq:x_tot}

%NOTE: Ref 49 is woodside2006direct
%-------------------------------------------------------------------------

\subsubsection{Computation of the Equilibrium FDC}

Let us consider the case where thermal fluctuations are neglected in the FDC computation. Thus, at a given value of $\lambda$, the system is always in the state of minimum energy, $\Delta G_{\text{eq}}(\lambda) = \Delta G_{\text{tot}}(\lambda,n^*)$.
To compute the equilibrium free energy of the system, let us first introduce the system partition function, $Z$. At each $\lambda$, this is defined as the sum over all the possible states, i.e., all the possible sequences of $n$ open base pairs, which is
\begin{equation}
    \label{eq:Z}
    Z(\lambda) = \sum_{n=0}^N \exp\left(-\frac{\Delta G_{\text{tot}}(\lambda,n)}{\kt}\right) \, ,
\end{equation}
where $N$ is the total number of base pairs of the sequence. Finally, by recalling that $\Delta G = -\kt \ln Z$, the equilibrium force is given by:
\begin{equation}
    \label{eq:f_theo}
    f_{\text{eq}}(\lambda) \equiv \frac{\partial \Delta G(x_{\text{eq}})}{\partial \lambda}  = -\kt \frac{\partial \ln Z(\lambda)}{\partial \lambda} .
\end{equation}
Computing Eq.\eqref{eq:Z} requires solving the transcendental Eq.\eqref{eq:x_tot} with respect to $f$ (that can be performed numerically) and then computing Eq.\eqref{eq:G_eq} for all $n \in [0,N-1]$. For each $\lambda$, the value $n^*$ minimizing the equilibrium free-energy $\Delta G_{\text{eq}} = \Delta G_{\text{tot}}(\lambda,n^*(\lambda))$ gives the most probable number of open base-pairs. Eventually, the computation of the equilibrium force in Eq.\eqref{eq:f_theo} gives a theoretical prediction for the unzipping curve of a given sequence (\tcr{Extended Data Fig. 5}). 

%-------------------------------------------------------------------------

\subsubsection{Equilibrium Free Energy}

The total free energy in Eq.\eqref{eq:G_eq} is the sum of two main contributions: the hybridization energy $\Delta G_0(n)$, which linearly depends on the number of hybridized NNBPs $n$, and the stretching energy $\Delta G_{\rm el}(\lambda, n) = \Delta G_\text{b}(x_\text{b}) + \Delta G_\text{h}(x_\text{h}) +\Delta G_\text{ss}(x_\text{ss},n) + \Delta G_\text{d}(x_\text{d})$ depending on both $n$ and $\lambda$. 
For a given $\lambda$, the equilibrium configuration of the system is that with minimum $\Delta G_{\rm el}(\lambda, n^*)$ and maximum $\Delta G_0(n^*)$ among all possible values of $n$. Notice that for a hairpin of $N$ bp, $n$ ranges from 0 (native state) to $N-1$ NNBPs (totally unfolded), which gives $N-1$ possible system configurations for each value of $\lambda$.

Let us suppose that the system starts at equilibrium, with $n_1$ open bp. Upon increasing $\lambda$, the elastic term in Eq.\eqref{eq:G_eq} also increases. The number of open bp, $n_1$, remains constant until a value of $n=n_2>n_1$ is found so that $\Delta G_{\text{tot}}(\lambda,n_1) \equiv \Delta G_{\text{tot}}(\lambda,n_2)$ (\tcr{Extended Data Fig. 4A, top}): even though the total energy of these two states is the same, the energetic internal balance is different (\tcr{Extended Data Fig. 4A, bottom}). 
The system minimizes the elastic free energy and switches to state $n_2$ by releasing $\Delta n = n_2 - n_1$ bp. Notice that, despite opening $\Delta n$ bp increases the system's energy, the released ssDNA causes the elastic contribution to decrease. In general, $\Delta G_{\text{el}} \gg \Delta G_{0}$ so the global balance of the state $n_2$ is lower than the one of $n_1$.
Therefore, the equilibrium free energy of hybridization, $\Delta G_0(n*)$, is a step function increasing with $\lambda$ (\tcr{Extended Data Fig. 4B}) with each discontinuity corresponding to a rip along the equilibrium FDC.

%------------------------------------------------------------------------
%************************************************************************
%------------------------------------------------------------------------

\subsection{Fit of the NNBP parameters}
\label{subsec:fit}

The $T$-dependent NNBP entropies and enthalpies permit us to derive the heat capacity changes $\Delta c_{p,i}$ for each motif from the relations,  
\begin{subequations}
	\label{eq:dSdHfit}
	\begin{align}
		\label{eq:dSfit}
		\Delta s_i &= \Delta s_{m,i} + \Delta c_{p,i} \log(T/T_{m,i}) \\
		\label{eq:dHfit}
		\Delta h_i &= \Delta h_{m,i} + \Delta c_{p,i} (T-T_{m,i}) \, ,
	\end{align}
\end{subequations}
where $T_{m,i}$ is the melting temperature of motif $i$, and $\Delta s_{m,i}$ and $\Delta h_{m,i}$ are the entropy and enthalpy at $T=T_{m,i}$, respectively.
The extraction of the NNBP thermodynamics parameters ($\Delta c_{p,i}, \Delta s_{i}, \Delta h_{i}, T_{m,i}$) has to be carried out carefully as the results are susceptible to experimental errors and parameters initialization. In particular, $\Delta s_{m,i}$, $\Delta h_{m,i}$, and $T_{r,i}$ strongly depend on their initialization values when directly fitted from Eqs.\eqref{eq:dSdHfit} as an error in $\Delta s_{m,i}$ ($\Delta h_{m,i}$) get compensated by $T_{m,i}$ and \textit{vice versa}.  

To derive the $\Delta c_{p,i}$, we fit the NNBP entropies to the equation $\Delta s_i(T) = A_i + \Delta c_{p,i} \log(T)$, being $A_i= \Delta s_{m,i} - \Delta c_{p,i} \log(T_{m,i})$. Notice that we derive $\Delta c_{p,i}$ from the NNBP entropies as they are obtained from the experimental data, in contrast to enthalpies that are computed from the free energies.
Given $\Delta c_{p,i}$, we fit the NNBP free energies, $\Delta g_i(T)$, to the equation 
\begin{equation}
	\label{eq:Tmelting}
	\begin{split}
		\Delta g_i(T) &= \Delta h_i(T) - T\Delta s_i(T) = \\
		 &= \Delta h_{m,i} + \Delta c_{p,i}(T - T_{m,i}) - T \left(\Delta s_{m,i} + \Delta c_{p,i} \log{\left(\frac{T}{T_{m,i}}\right)} \right) \\
		 &= B_i + \Delta c_{p,i}T - T \left(A_i + \Delta c_{p,i} \log{(T)} \right) \, .
	\end{split}
\end{equation}
obtained by combining Eqs.\eqref{eq:dSdHfit} (blue dashed lines in Fig. 3B, main text). Notice that $B_i = \Delta h_{m,i} - \Delta c_{p,i}T_{m,i}$.
By definition, $T_{m,i}$ is the high temperature value where $\Delta g_i(T_{m,i})=0$. 
Finally, a new fit to Eqs.\eqref{eq:dSfit} and \eqref{eq:dHfit} by using the previously derived values of $\Delta c_{p,i}$ and $T_{m,i}$ (red and blue dashed lines in \tcr{Fig. 2D}, main text), gives $\Delta s_{m,i}$ and $\Delta h_{m,i}$. The results are shown in \tcr{Fig. 4 and Table 1 of the main text}.

% =====================================================================
% *********************************************************************
% =====================================================================

\clearpage

\renewcommand{\thesection}{Extended Data: Figures}% Remove section references...
\section{}

%FROM FDC to FEC
\vfill
\begin{figure*}[h]
    \centering
    \includegraphics[width=\textwidth]{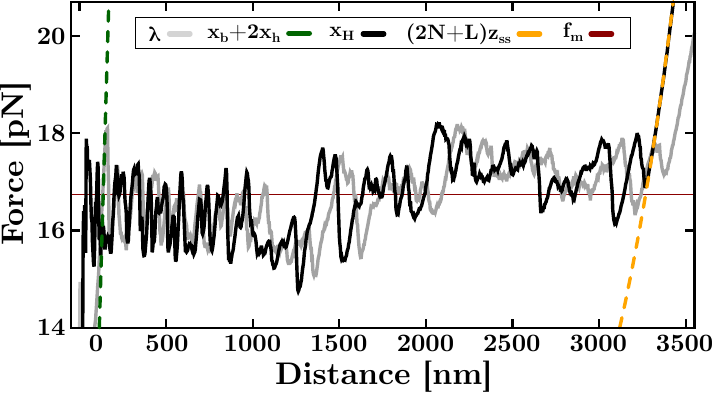}
    \caption{\textbf{Computation of the FEC from the experimental FDC}. The force versus hairpin extension, $x_H$, (black line) is computed by subtracting to the trap position, $\lambda$, (grey line) the elastic contribution of the optically trapped bead, $x_{b}$, and DNA handles, $2x_{h}$, (green dashed line). 
    To measure the $T$-dependent ssDNA elasticity, we fit the FEC after the last rip to the WLC model (orange dashed line). Notice that the average unzipping force, $f_m$, (red line) remains constant upon computing the FEC. Data are shown at $T=25\deg$C.}
    \label{figS:FDCtoFEC}
\end{figure*}
\vfill
\clearpage
\newpage

%FEC vs T
\null
\vfill
\begin{figure*}[h]
    \centering
    \includegraphics[width=0.9\textwidth]{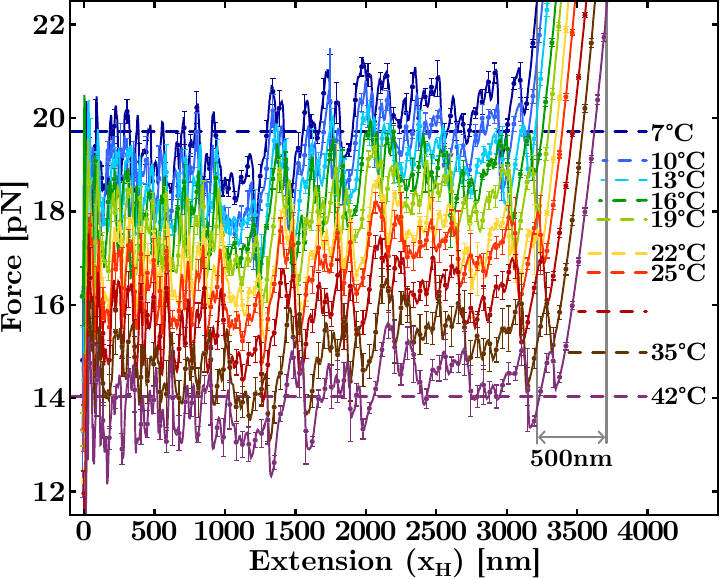}
    \caption{\textbf{T-dependence of the measured force versus hairpin extension.} At each $T$, average unzipping forces are shown by dashed lines. The extension change over the studied temperature range is $\sim 500$nm (grey vertical lines).}
    \label{figS:FECvsT}
\end{figure*}
\vfill
\clearpage
\newpage

%CLAUSIUS-CLAPEYRON EQUATION APPLIED TO THE FULL FDC
\null
\vfill
\begin{figure*}[h]
    \centering
    \includegraphics[width=\textwidth]{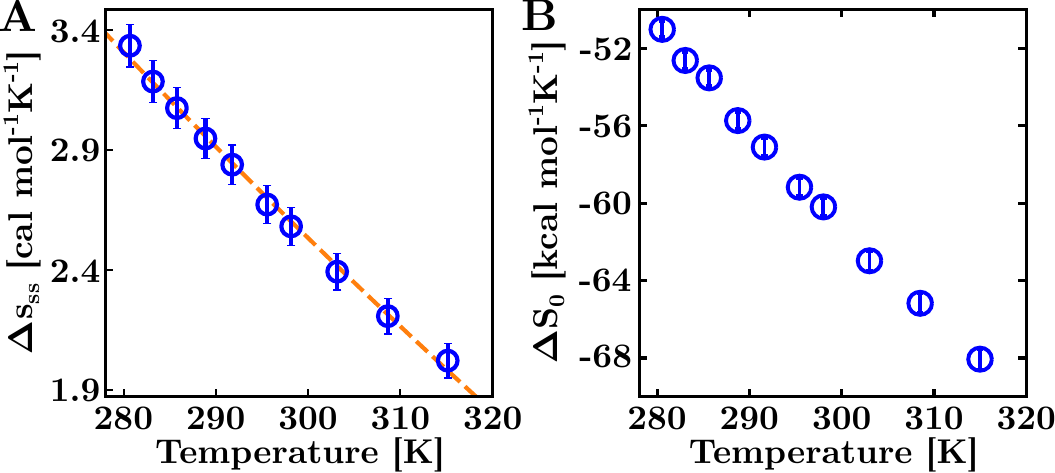}
    \caption{\textbf{Clausius-Clapeyron equation applied over the full FDC.} (\textbf{A}) $T$-dependence of the entropy change per base, $\Delta s_{ss}$. It accounts for the work to stretch the ssDNA and orient the folded molecule along the direction of the external force, from $f=0$pN to $f_{\rm m}(T)$ (integral term in \tcr{Eq.(2)}, main text). The results are reported in Extended Data Table \ref{tabS:Tdependence}. A fit to data according to $\Delta s_{\rm ss}(T) = \Delta s_{\rm ss, 0} + \Delta c^{\rm ss}_{p} \log(T/T_{m})$ (orange dashed line), gives the ssDNA heat capacity change per base at zero force, $\Delta c^{\rm ss}_{p} = -11.2 \pm 0.2 \rm \, cal \, mol^{-1}K^{-1}$. (\textbf{B}) $T$-dependence of the total entropy change, $\Delta S_0(T)$, upon unzipping the 3.6kbp DNA hairpin measured using the Clausius-Clapeyron equation (see \tcr{Eq.(2)}, main text).}
    \label{figS:CCfullFDC}
\end{figure*}
\vfill
\clearpage
\newpage

%THEORETICAL FDC
\null
\vfill
\begin{figure*}[h]
    \centering
    \includegraphics[width=\textwidth]{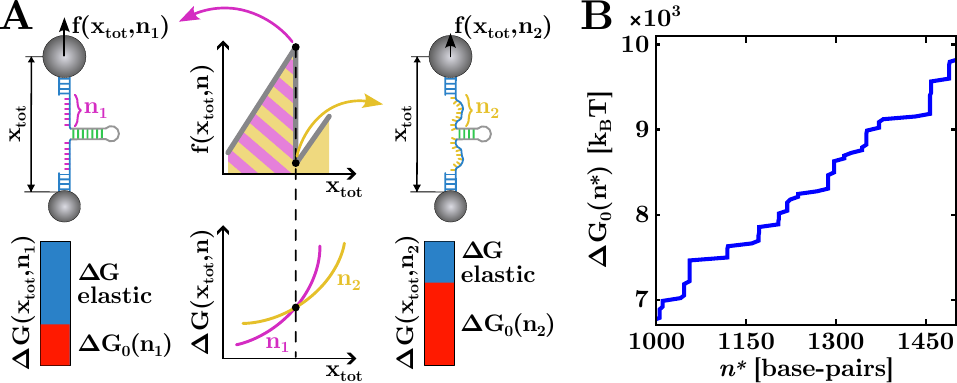}
    \caption{\textbf{Derivation of the theoretical FDC.} (\textbf{A}) Schematics of the stretching and hybridization energy contributions. Upon unzipping, the molecule has $n_1$ open bp before the force rip (left) and $n_2>n_1$ open bp after the rip (right). At the force rip (black dots), the total free energy of the system is the same in both states, and the system changes from the highest free energy branch ($n_1$, pink) to the lowest energy branch ($n_2$, pink). (\textbf{B}) The free energy of hybridization upon unzipping the hairpin is a monotonically increasing step function, with each discontinuity corresponding to a rip along the equilibrium FDC.}
    \label{figS:FDCtheo}
\end{figure*}
\vfill
\clearpage
\newpage

%THEORETICAL FDCs vs T
\null
\vfill
\begin{figure*}[h]
    \centering
    \includegraphics[width=\textwidth]{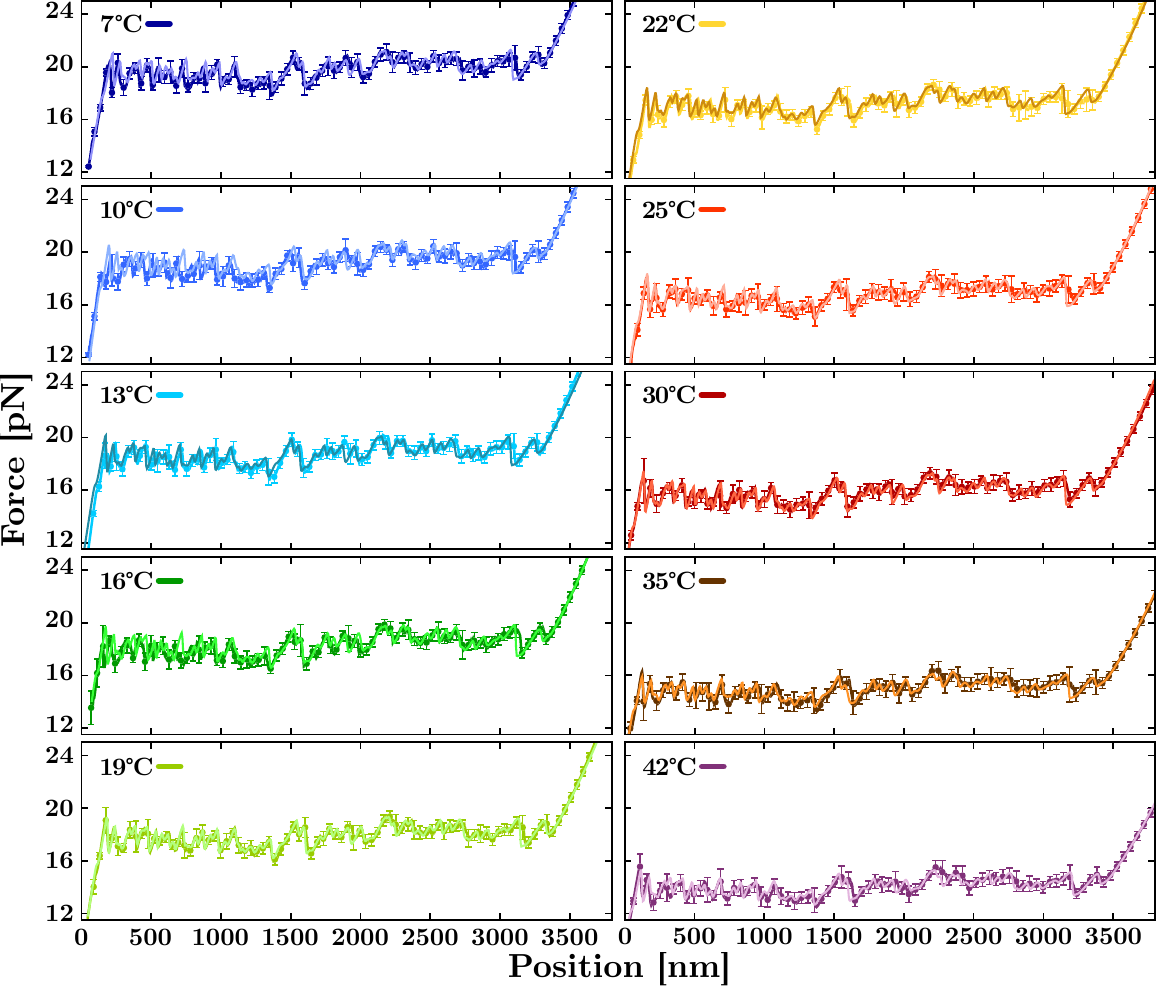}
    \caption{\textbf{$\mathbf{T}$-dependent theoretical FDC predictions.} Experimental FDCs (dark-colored solid lines and points) and theoretical predictions (light-colored lines) obtained with the free energy parameters derived at each $T$. Error bars indicate the variability of the experimental FDCs.}
    \label{figS:theoFDCvsT}
\end{figure*}
\vfill
\clearpage
\newpage

\iffalse
%NNBP UNZ vs LITERATURE
\null
\vfill
\begin{figure*}[h]
    \centering
    \includegraphics[width=1.\textwidth]{FigureS_NNBPthermoComparison.pdf}
    \caption{\textbf{NNBP thermodynamics at the melting temperature.} The NNBP entropies (\textbf{A}) and enthalpies (\textbf{B}) values (vertical axis) at the melting temperature (cf. \tcr{Table 1, main text}) are compared to the values of the unified oligonucleotide set \tcr{$^{Ref\#}$} (red circles), Huguet \textit{et al.}, 2010 \tcr{$^{Ref\#}$} (blue squares), and Huguet \textit{et al.}, 2017 \tcr{$^{Ref\#}$} (orange triangles) plotted on the horizontal axis. The perfect agreement would imply all data points falling on the dashed grey line $x=y$. Deviations with respect to the results by Huguet \textit{et al.} are due to those measurements being carried out at room temperature and considered as $T$-independent.}
    \label{figS:NNBPcomp}
\end{figure*}
\vfill
\clearpage
\newpage
\fi

%TM UNZ vs LITERATURE
\null
\vfill
\begin{figure*}[h]
    \centering
    \includegraphics[width=\textwidth]{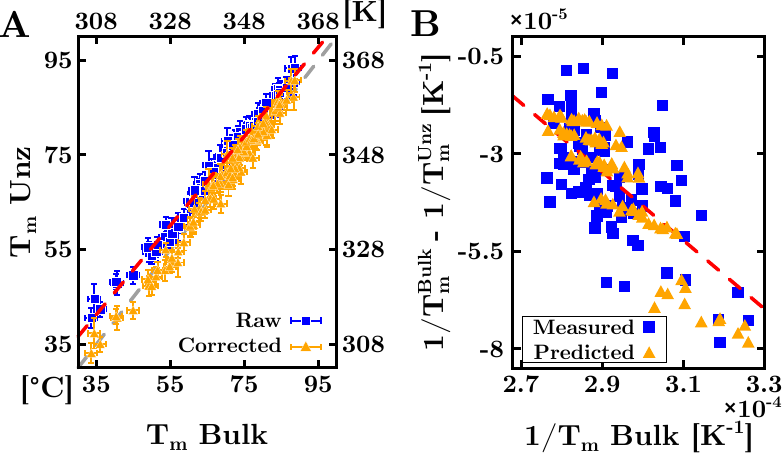}
    \caption{\textbf{Prediction of the DNA duplexes melting temperatures.} (\textbf{A}) Comparison of the melting temperatures for the set of 92 DNA oligos studied by Owczarzy \textit{et al.} in Ref.\tcb{$\,^{55}$} (horizontal axis) and the values predicted with the unzipping energy parameters (vertical axis). Perfect agreement between the two data sets would imply all points falling on the dashed grey line $x=y$. Predictions obtained with \tcr{Eq.(10)} of \tcr{Sec. 6, Methods} (blue squares) show a systematic discrepancy with respect to the experimental values (dashed red line). By accounting for the entropic correction (\tcr{Eq.(11) of Sec. 6, Methods}), predictions agree with the experimental measurements within errors (orange triangles). Results are reported in Extended Data Table \ref{tabS:Tmelting}. (\textbf{B}) Derivation of the entropic correction, $\delta \Delta s$. To do this, we subtracted the inverse of the measured, $T^{Bulk}_m$, and predicted, $T^{Unz}_m$, melting temperatures (blue squares). This equals the difference between the inverse of \tcr{Eq.(11) and Eq.(10)} (\tcr{see Eq.(12) in Sec. 6, Methods}). A linear fit to data (dashed red line) yields $\delta\Delta s= 6(1) \, {\rm cal \, mol^{-1}K^{-1}} \sim 4R\log 2$, where $R=1.987 \, {\rm cal \, mol^{-1}K^{-1}}$ is the ideal gas constant. The orange triangles show the theoretical correction to $T_m$ per DNA duplex predicted by assuming $\delta \Delta s \equiv 4R\log 2$.}
    \label{figS:Tmelting}
\end{figure*}
\vfill
\clearpage
\newpage

%NOTE: Ref 55 is owczarcy2008predicting

%COLD DENATURATION
\null
\vfill
\begin{figure*}[h]
    \centering
    \includegraphics[width=0.8\textwidth]{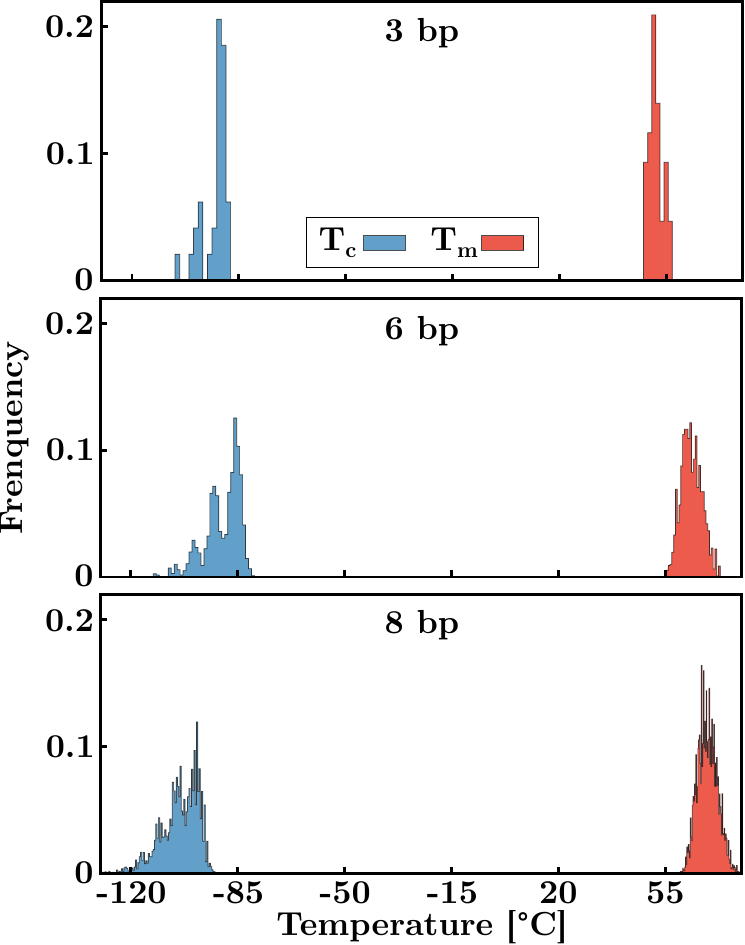}
    \caption{\textbf{Prediction of DNA cold denaturation temperatures.} Histograms of the melting (red) cold denaturation (blue) temperatures predicted using the ten NNBP thermodynamics parameters (\tcr{Table 1, main}) for all possible DNA sequences of 3, 6, and 8 bp ending with a GAAA tetraloop.}
    \label{figS:ColdDen}
\end{figure*}
\vfill
\clearpage
\newpage

\clearpage

% ********************************************************************************
% --------------------------------------------------------------------------------
% ********************************************************************************

\renewcommand{\thesection}{Extended Data: Tables}% Remove section references...
\section{}

%T-DEPENDENT FDC PROPERTIES
\vfill
\begin{table}[h]
\centering
\caption{\label{tabS:Tdependence} \textbf{T-dependence of the DNA FDCs}}
\begin{adjustbox}{width=\textwidth}
\begin{tabular}{cc|cccc}
$\mathbf{T}$ \textbf{[$^{\circ}$C]} & $\mathbf{T}$ \textbf{[K]} & $\mathbf{f_{\rm m}}$ \textbf{[pN]} & $\mathbf{l_p}$ \textbf{[nm]} & $\mathbf{d_b}$ \textbf{[nm]} & $\mathbf{\Delta s_{ss}}$ \textbf{[cal mol$^{-1}$K$^{-1}$]}\\
\toprule
$7$ & $280$ & $19.72$ $(3)$ & $0.74$ $(7)$ & $0.647$ $(3)$ & 3.33 (2) \\
$10$ & $283$ & $19.08$ $(4)$ & $0.68$ $(2)$ & $0.631$ $(9)$ & 3.18 (2) \\
$13$ & $286$ & $18.71$ $(4)$ & $0.73$ $(3)$ & $0.662$ $(1)$ & 3.07 (2) \\
$16$ & $289$ & $18.26$ $(7)$ & $0.67$ $(2)$ & $0.672$ $(1)$ & 2.95 (2) \\
$19$ & $292$ & $17.87$ $(4)$ & $0.78$ $(2)$ & $0.655$ $(1)$ & 2.84 (2) \\
$22$ & $295$ & $17.12$ $(2)$ & $0.79$ $(3)$ & $0.657$ $(1)$ & 2.67 (2) \\
$25$ & $298$ & $16.75$ $(3)$ & $0.77$ $(2)$ & $0.647$ $(1)$ & 2.58 (2) \\
$30$ & $303$ & $15.86$ $(2)$ & $0.75$ $(2)$ & $0.665$ $(1)$ & 2.39 (2) \\
$35$ & $308$ & $14.96$ $(3)$ & $0.88$ $(2)$ & $0.639$ $(1)$ & 2.21 (1) \\
$42$ & $315$ & $14.06$ $(4)$ & $0.88$ $(4)$ & $0.641$ $(1)$ & 2.02 (1) \\
\bottomrule
\end{tabular}
\end{adjustbox}
\caption*{FDC average unzipping force, $f_{\rm m}$, persistence length, $l_p$, interphosphate distance, $d_b$, and ssDNA stretching entropy per base, $\Delta s_{ss}$ in the studied temperature range (in Celsius and Kelvin degrees). The errors (in brackets) refer to the last digit. The error in temperature is $\pm 1^{\circ}$C (K).}
\end{table}
\vfill
\clearpage
\newpage

%T-DEPENDENT NNBP ENTROPIES
\vfill
{\renewcommand{\arraystretch}{1.5}
\begin{sidewaystable}[h]
\centering
\caption{NNBP $\Delta s_{0,i}$ [$\rm cal \, mol^{-1} K^{-1}$] at different temperatures.}
\begin{adjustbox}{width=\textwidth}
\begin{tabular}{c|cccccccccc}
\toprule
\textbf{Temperature $\pm 1$ [K]}  & $\mathbf{280}$ &  $\mathbf{283}$ &  $\mathbf{285}$ &  $\mathbf{288}$ &  $\mathbf{291}$ &  $\mathbf{295}$ &  $\mathbf{298}$ &  $\mathbf{303}$ & $\mathbf{308}$ &  $\mathbf{315}$ \\
\midrule
\textbf{AA/TT} & -12.4 (5) & -12.6 (4) & -15.7 (7) & -12.8 (3) & -15.0 (5) & -16.0 (5) & -16.3 (6) & -16.9 (5) & -16.0 (3) & -18.3 (4) \\
\textbf{AC/TG} & -15.4 (2) & -15.6 (1) & -16.4 (2) & -16.4 (1) & -16.9 (2) & -17.5 (1) & -17.7 (1) & -18.3 (1) & -18.6 (1) & -19.6 (2) \\
\textbf{AG/TC} & -12.0 (3) & -13.1 (2) & -11.8 (4) & -14.3 (2) & -13.3 (4) & -13.8 (4) & -14.5 (3) & -15.0 (4) & -16.2 (3) & -16.6 (3) \\
\textbf{AT/TA} & -15.1 (1) & -15.3 (1) & -16.1 (1) & -15.6 (1) & -16.4 (1) & -17.1 (2) & -16.7 (1) & -17.3 (1) & -17.3 (1) & -17.9 (1) \\
\textbf{CA/GT} & -12.8 (2) & -13.7 (2) & -13.8 (2) & -14.9 (2) & -15.1 (2) & -16.6 (4) & -16.0 (2) & -17.3 (3) & -19.4 (7) & -18.4 (2) \\
\textbf{CC/GG} & -12.4 (3) & -13.3 (3) & -12.2 (4) & -14.6 (2) & -13.5 (4) & -13.7 (4) & -14.8 (3) & -15.2 (3) & -16.0 (3) & -16.4 (3) \\
\textbf{CG/GC} & -22.2 (3) & -22.4 (4) & -21.4 (5) & -23.8 (2) & -22.8 (4) & -22.6 (4) & -23.7 (3) & -24.1 (3) & -24.2 (4) & -24.6 (3) \\
\textbf{GA/CT} & -12.4 (4) & -13.4 (3) & -12.4 (5) & -14.6 (3) & -14.1 (4) & -14.5 (4) & -14.7 (5) & -15.2 (6) & -16.6 (4) & -17.3 (3) \\
\midrule
\textbf{GC/CG} & -20.8 (3) & -21.3 (3) & -19.7 (5) & -22.8 (2) & -21.5 (4) & -21.8 (3) & -22.7 (3) & -23.5 (2) & -24.4 (3) & -23.7 (3) \\
\textbf{TA/AT} & -16.1 (1) & -16.1 (1) & -17.1 (2) & -16.2 (1) & -16.9 (1) & -17.2 (1) & -17.5 (2) & -17.7 (2) & -16.7 (3) & -18.2 (1) \\
\bottomrule
\end{tabular}
\end{adjustbox}
\caption*{\label{tabS:NNBPdSvsT} The 10 DNA entropies measured from unzipping a 3.6kbp hairpin in the temperature range $[280,315]$ K (see text). The entropy of the last two motifs (GC/CG and TA/AT) has been computed by applying the circular symmetry relations. The error (in brackets) refers to the last digit.}
\end{sidewaystable}}
\vfill
\clearpage
\newpage

%T-DEPENDENT NNBP FREE-ENERGIES
\vfill
{\renewcommand{\arraystretch}{1.5}
\begin{sidewaystable}[h]
\centering
\caption{NNBP $\Delta g_{0,i}$ [$\rm kcal \, mol^{-1}$] at different temperatures.}
\begin{adjustbox}{width=\textwidth}
\begin{tabular}{c|cccccccccc}
\toprule
\textbf{Temperature $\pm 1$ [K]}  & $\mathbf{280}$ &  $\mathbf{283}$ &  $\mathbf{285}$ &  $\mathbf{288}$ &  $\mathbf{291}$ &  $\mathbf{295}$ &  $\mathbf{298}$ &  $\mathbf{303}$ & $\mathbf{308}$ &  $\mathbf{315}$ \\
\midrule
\textbf{AA/TT} & $-1.57 (2)$ & $-1.49 (2)$ & $-1.55 (2)$ & $-1.47 (2)$ & $-1.43 (1)$ & $-1.43 (1)$ & $-1.30 (1)$ & $-1.27 (1)$ & $-1.19 (1)$ & $-1.15 (1)$ \\
\textbf{AC/TG} & $-1.52 (2)$ & $-1.38 (1)$ & $-1.53 (2)$ & $-1.35 (1)$ & $-1.47 (2)$ & $-1.34 (1)$ & $-1.43 (1)$ & $-1.36 (1)$ & $-1.17 (1)$ & $-1.21 (1)$ \\
\textbf{AG/TC} & $-1.73 (2)$ & $-1.66 (2)$ & $-1.49 (2)$ & $-1.60 (2)$ & $-1.54 (2)$ & $-1.42 (1)$ & $-1.41 (1)$ & $-1.25 (1)$ & $-1.35 (1)$ & $-1.25 (1)$ \\
\textbf{AT/TA} & $-1.37 (1)$ & $-1.43 (1)$ & $-1.27 (1)$ & $-1.32 (1)$ & $-1.25 (1)$ & $-1.28 (1)$ & $-1.17 (1)$ & $-1.00 (1)$ & $-1.19 (1)$ & $-1.00 (1)$ \\
\textbf{CA/GT} & $-1.86 (2)$ & $-1.92 (2)$ & $-1.82 (2)$ & $-1.88 (2)$ & $-1.82 (2)$ & $-1.91 (2)$ & $-1.65 (2)$ & $-1.68 (2)$ & $-1.70 (2)$ & $-1.52 (2)$ \\
\textbf{CC/GG} & $-2.03 (2)$ & $-1.86 (2)$ & $-2.04 (2)$ & $-1.91 (2)$ & $-2.00 (2)$ & $-1.88 (2)$ & $-1.91 (2)$ & $-1.86 (2)$ & $-1.56 (2)$ & $-1.70 (2)$ \\
\textbf{CG/GC} & $-2.51 (3)$ & $-2.52 (3)$ & $-2.39 (2)$ & $-2.46 (3)$ & $-2.35 (2)$ & $-2.24 (2)$ & $-2.43 (2)$ & $-2.30 (2)$ & $-2.03 (2)$ & $-1.93 (2)$ \\
\textbf{GA/CT} & $-1.52 (2)$ & $-1.59 (2)$ & $-1.55 (2)$ & $-1.48 (2)$ & $-1.49 (2)$ & $-1.42 (1)$ & $-1.52 (2)$ & $-1.46 (2)$ & $-1.26 (1)$ & $-1.20 (1)$ \\
\midrule
\textbf{GC/CG} & $-2.79 (3)$ & $-2.83 (3)$ & $-2.51 (3)$ & $-2.78 (3)$ & $-2.55 (3)$ & $-2.53 (3)$ & $-2.49 (3)$ & $-2.36 (2)$ & $-2.34 (2)$ & $-2.11 (2)$ \\
\textbf{TA/AT} & $-1.31 (1)$ & $-1.19 (1)$ & $-1.10 (1)$ & $-1.11 (1)$ & $-1.10 (1)$ & $-1.00 (1)$ & $-1.00 (1)$ & $-0.74 (1)$ & $-0.97 (1)$ & $-0.88 (1)$ \\
\bottomrule
\end{tabular}
\end{adjustbox}
\caption*{\label{tabS:NNBPdGvsT} The 10 DNA free-energies measured from unzipping a 3.6kbp hairpin in the temperature range $[280,315]$ K (see text). The entropy of the last two motifs (GC/CG and TA/AT) has been computed by the applying circular symmetry relations. The error (in brackets) refers to the last digit.}
\end{sidewaystable}}
\vfill
\clearpage
\newpage

%T-DEPENDENT NNBP ENTHALPIES
\vfill
{\renewcommand{\arraystretch}{1.5}
\begin{sidewaystable}[h]
\centering
\caption{NNBP $\Delta h_{0,i}$ [$\rm kcal \, mol^{-1}$] at different temperatures.}
\begin{adjustbox}{width=\textwidth}
\begin{tabular}{c|cccccccccc}
\toprule
\textbf{Temperature $\pm 1$ [K]}  & $\mathbf{280}$ &  $\mathbf{283}$ &  $\mathbf{285}$ &  $\mathbf{288}$ &  $\mathbf{291}$ &  $\mathbf{295}$ &  $\mathbf{298}$ &  $\mathbf{303}$ & $\mathbf{308}$ &  $\mathbf{315}$ \\
\midrule
\textbf{AA/TT} & -5.05 (15) & -5.04 (12) & -6.04 (21) & -5.16 (08) & -5.83 (16) & -6.17 (15) & -6.16 (17) & -6.39 (15) & -6.13 (09) & -6.93 (14) \\
\textbf{AC/TG} & -5.84 (05) & -5.80 (03) & -6.21 (05) & -6.10 (04) & -6.41 (05) & -6.51 (04) & -6.71 (05) & -6.91 (04) & -6.90 (03) & -7.38 (06) \\
\textbf{AG/TC} & -5.09 (10) & -5.36 (07) & -4.86 (13) & -5.74 (06) & -5.41 (12) & -5.51 (12) & -5.73 (10) & -5.80 (12) & -6.34 (09) & -6.49 (09) \\
\textbf{AT/TA} & -5.60 (03) & -5.76 (03) & -5.88 (04) & -5.82 (03) & -6.03 (04) & -6.32 (05) & -6.15 (03) & -6.26 (04) & -6.54 (04) & -6.65 (03) \\
\textbf{CA/GT} & -5.45 (05) & -5.81 (05) & -5.76 (06) & -6.19 (06) & -6.22 (07) & -6.81 (11) & -6.41 (05) & -6.93 (09) & -7.67 (20) & -7.31 (06) \\
\textbf{CC/GG} & -5.51 (10) & -5.62 (08) & -5.52 (12) & -6.13 (06) & -5.93 (11) & -5.93 (12) & -6.32 (08) & -6.46 (10) & -6.51 (10) & -6.88 (08) \\
\textbf{CG/GC} & -8.75 (09) & -8.86 (11) & -8.49 (14) & -9.32 (07) & -9.01 (11) & -8.92 (13) & -9.51 (09) & -9.60 (09) & -9.50 (12) & -9.68 (10) \\
\textbf{GA/CT} & -4.99 (11) & -5.37 (08) & -5.09 (14) & -5.71 (08) & -5.59 (11) & -5.70 (13) & -5.91 (14) & -6.07 (17) & -6.38 (12) & -6.63 (11) \\
\midrule
\textbf{GC/CG} & -8.61 (10) & -8.85 (10) & -8.15 (14) & -9.38 (06) & -8.82 (11) & -8.98 (11) & -9.27 (09) & -9.48 (07) & -9.87 (09) & -9.57 (11) \\
\textbf{TA/AT} & -5.84 (04) & -5.75 (03) & -5.99 (05) & -5.79 (03) & -6.04 (04) & -6.07 (04) & -6.21 (05) & -6.11 (05) & -6.14 (08) & -6.61 (05) \\
\bottomrule
\end{tabular}
\end{adjustbox}
\caption*{\label{tabS:NNBPdHvsT} The 10 DNA enthalpies measured from unzipping a 3.6kbp hairpin in the temperature range $[280,315]$ K (see text). The entropy of the last two motifs (GC/CG and TA/AT) has been computed by the applying circular symmetry relations. The error (in brackets) refers to the last digit.}
\end{sidewaystable}}
\vfill
\clearpage
\newpage

%PREDICTION OF OLIGOS MELTING T
\vfill
\setlength\tabcolsep{5.5pt}
\setlength\LTleft{0pt}
\setlength\LTright{0pt}
\footnotesize
\begin{longtable}[c]{l | ccccc}
\caption{DNA Oligos Melting Temperatures [$^{\circ}$C]}\\
\toprule
\textbf{Sequence ($5^{\prime} \rightarrow 3^{\prime}$)} & $\mathbf{T^{\rm \textbf{Exp}}}$ & \textbf{T$\mathbf{_{Bi}^{Unz}}$} & \textbf{T$\mathbf{_{Uni}^{Unz}}$} & \textbf{T$\mathbf{^{UO}}$} & \textbf{T$\mathbf{^{Hug}}$} \\ \\
\hline
\endhead
\hline
\endfoot
\endlastfoot
ATCAATCATA & 33.6 & 40.6 & 33.1 & 34.0 & 32.3 \\
TTGTAGTCAT & 36.0 & 42.4 & 35.0 & 36.7 & 36.2 \\
GAAATGAAAG & 34.4 & 44.6 & 37.3 & 34.6 & 33.8 \\
CCAACTTCTT & 40.6 & 47.9 & 40.6 & 40.4 & 38.9 \\
ATCGTCTGGA & 44.9 & 49.6 & 42.1 & 46.2 & 45.3 \\
AGCGTAAGTC & 40.3 & 48.2 & 40.9 & 45.1 & 43.4 \\
CGATCTGCGA & 49.1 & 54.5 & 47.3 & 50.5 & 50.2 \\
TGGCGAGCAC & 55.3 & 59.5 & 52.4 & 56.3 & 54.7 \\
GATGCGCTCG & 53.5 & 57.7 & 50.6 & 54.0 & 52.9 \\
GGGACCGCCT & 57.0 & 59.8 & 52.3 & 58.6 & 55.3 \\
CGTACACATGC & 49.9 & 53.8 & 47.3 & 51.2 & 50.6 \\
CCATTGCTACC & 48.9 & 55.5 & 48.7 & 49.6 & 48.2 \\
TACTAACATTAACTA & 51.1 & 54.6 & 49.3 & 51.9 & 50.1 \\
ATACTTACTGATTAG & 51.5 & 56.0 & 50.6 & 49.7 & 50.4 \\
GTACACTGTCTTATA & 54.8 & 56.7 & 51.4 & 54.8 & 54.2 \\
GTATGAGAGACTTTA & 55.4 & 58.5 & 53.2 & 54.8 & 54.2 \\
TTCTACCTATGTGAT & 53.7 & 60.3 & 55.0 & 55.1 & 55.7 \\
AGTAGTAATCACACC & 57.1 & 59.8 & 54.5 & 56.9 & 57.2 \\
ATCGTCTCGGTATAA & 58.6 & 61.8 & 56.5 & 58.9 & 59.9 \\
ACGACAGGTTTACCA & 61.3 & 65.3 & 60.1 & 63.6 & 63.6 \\
CTTTCATGTCCGCAT & 62.8 & 68.0 & 62.9 & 63.0 & 62.6 \\
TGGATGTGTGAACAC & 60.4 & 64.8 & 59.8 & 62.3 & 62.1 \\
ACCCCGCAATACATG & 62.9 & 68.9 & 63.8 & 65.6 & 64.5 \\
GCAGTGGATGTGAGA & 63.3 & 68.1 & 63.0 & 64.6 & 64.2 \\
GGTCCTTACTTGGTG & 60.3 & 65.2 & 60.0 & 62.0 & 61.7 \\
CGCCTCATGCTCATC & 65.8 & 70.9 & 65.8 & 66.5 & 65.9 \\
AAATAGCCGGGCCGC & 70.4 & 75.8 & 70.7 & 72.7 & 70.9 \\
CCAGCCAGTCTCTCC & 66.7 & 70.9 & 65.7 & 67.7 & 66.7 \\
GACGACAAGACCGCG & 68.6 & 69.7 & 64.7 & 69.7 & 70.3 \\
CAGCCTCGTCGCAGC & 72.0 & 74.8 & 69.8 & 73.0 & 72.7 \\
CTCGCGGTCGAAGCG & 70.7 & 73.7 & 68.7 & 72.9 & 73.6 \\
GCGTCGGTCCGGGCT & 74.1 & 76.5 & 71.4 & 77.8 & 76.1 \\
TATGTATATTTTGTAATCAG & 61.2 & 64.9 & 60.8 & 58.6 & 59.5 \\
TTCAAGTTAAACATTCTATC & 61.5 & 67.6 & 63.6 & 60.6 & 62.6 \\
TGATTCTACCTATGTGATTT & 64.4 & 69.5 & 65.4 & 63.7 & 64.8 \\
GAGATTGTTTCCCTTTCAAA & 65.3 & 72.8 & 68.8 & 66.3 & 67.1 \\
ATGCAATGCTACATATTCGC & 68.9 & 74.7 & 70.8 & 69.2 & 69.6 \\
CCACTATACCATCTATGTAC & 64.4 & 67.5 & 63.4 & 63.9 & 65.2 \\
CCATCATTGTGTCTACCTCA & 68.5 & 73.0 & 69.0 & 69.4 & 69.7 \\
CGGGACCAACTAAAGGAAAT & 68.5 & 74.5 & 70.5 & 70.3 & 70.5 \\
TAGTGGCGATTAGATTCTGC & 71.2 & 74.6 & 70.6 & 71.1 & 70.9 \\
AGCTGCAGTGGATGTGAGAA & 73.1 & 78.0 & 74.1 & 74.5 & 74.0 \\
TACTTCCAGTGCTCAGCGTA & 73.6 & 76.2 & 72.2 & 76.0 & 74.6 \\
CAGTGAGACAGCAATGGTCG & 72.5 & 76.0 & 72.1 & 73.5 & 73.9 \\
CGAGCTTATCCCTATCCCTC & 70.3 & 75.5 & 71.4 & 71.3 & 71.9 \\
CGTACTAGCGTTGGTCATGG & 71.1 & 74.6 & 70.7 & 72.9 & 74.0 \\
AAGGCGAGTCAGGCTCAGTG & 76.3 & 79.4 & 75.5 & 77.2 & 77.0 \\
ACCGACGACGCTGATCCGAT & 77.3 & 78.3 & 74.4 & 78.7 & 78.6 \\
AGCAGTCCGCCACACCCTGA & 78.5 & 82.0 & 78.1 & 81.6 & 79.7 \\
CAGCCTCGTTCGCACAGCCC & 78.1 & 82.1 & 78.3 & 81.1 & 80.4 \\
GTGGTGGGCCGTGCGCTCTG & 81.0 & 83.2 & 79.4 & 83.6 & 82.0 \\
GTCCACGCCCGGTGCGACGG & 81.1 & 83.0 & 79.1 & 85.4 & 84.2 \\
GATATAGCAAAATTCTAAGTTAATA & 66.1 & 71.5 & 68.2 & 64.2 & 65.9 \\
ATAACTTTACGTGTGTGACCTATTA & 71.8 & 73.5 & 70.2 & 71.2 & 72.4 \\
GTTCTATACTCTTGAAGTTGATTAC & 67.7 & 72.2 & 68.9 & 67.3 & 69.9 \\
CCCTGCACTTTAACTGAATTGTTTA & 72.5 & 77.7 & 74.5 & 73.4 & 73.5 \\
TAACCATACTGAATACCTTTTGACG & 71.3 & 75.4 & 72.1 & 72.2 & 73.0 \\
TCCACACGGTAGTAAAATTAGGCTT & 73.8 & 78.2 & 74.9 & 74.6 & 75.8 \\
TTCCAAAAGGAGTTATGAGTTGCGA & 73.8 & 79.8 & 76.6 & 75.2 & 76.2 \\
AATATCTCTCATGCGCCAAGCTACA & 76.5 & 81.3 & 78.1 & 76.7 & 77.3 \\
TAGTATATCGCAGCATCATACAGGC & 75.0 & 78.8 & 75.5 & 75.5 & 75.8 \\
TGGATTCTACTCAACCTTAGTCTGG & 73.6 & 77.8 & 74.5 & 73.9 & 75.2 \\
CGGAATCCATGTTACTTCGGCTATC & 74.8 & 79.0 & 75.7 & 75.5 & 76.6 \\
CTGGTCTGGATCTGAGAACTTCAGG & 75.6 & 80.1 & 76.8 & 76.6 & 77.3 \\
ACAGCGAATGGACCTACGTGGCCTT & 81.0 & 83.6 & 80.4 & 82.7 & 82.1 \\
AGCAAGTCGAGCAGGGCCTACGTTT & 81.5 & 84.5 & 81.3 & 82.8 & 82.8 \\
GCGAGCGACAGGTTACTTGGCTGAT & 80.1 & 83.1 & 79.9 & 81.3 & 81.7 \\
AAAGGTGTCGCGGAGAGTCGTGCTG & 82.4 & 83.5 & 80.4 & 83.0 & 83.6 \\
ATGGGTGGGAGCCTCGGTAGCAGCC & 83.4 & 87.4 & 84.1 & 86.6 & 84.6 \\
CAGTGGGCTCCTGGGCGTGCTGGTC & 83.4 & 87.6 & 84.4 & 87.5 & 85.7 \\
GCCAACTCCGTCGCCGTTCGTGCGC & 84.6 & 86.9 & 83.8 & 88.1 & 88.0 \\
ACGGGTCCCCGCACCGCACCGCCAG & 88.3 & 90.4 & 87.2 & 93.0 & 90.1 \\
TTATGTATTAAGTTATATAGTAGTAGTAGT & 66.6 & 71.4 & 68.5 & 65.8 & 69.7 \\
ATTGATATCCTTTTCTATTCATCTTTCATT & 70.4 & 78.0 & 75.2 & 70.3 & 71.8 \\
AAAGTACATCAACATAGAGAATTGCATTTC & 73.2 & 78.8 & 76.1 & 73.0 & 74.6 \\
CTTAAGATATGAGAACTTCAACTAATGTGT & 71.8 & 77.1 & 74.3 & 71.8 & 74.2 \\
CTCAACTTGCGGTAAATAAATCGCTTAATC & 75.5 & 80.5 & 77.8 & 75.2 & 77.3 \\
TATTGAGAACAAGTGTCCGATTAGCAGAAA & 76.4 & 81.2 & 78.5 & 77.5 & 78.4 \\
GTCATACGACTGAGTGCAACATTGTTCAAA & 76.9 & 80.8 & 78.2 & 78.0 & 79.3 \\
AACCTGCAACATGGAGTTTTTGTCTCATGC & 78.7 & 83.7 & 81.1 & 80.3 & 80.1 \\
CCGTGCGGTGTGTACGTTTTATTCATCATA & 77.6 & 81.2 & 78.5 & 80.0 & 80.5 \\
GTTCACGTCCGAAAGCTCGAAAAAGGATAC & 78.7 & 82.1 & 79.4 & 79.5 & 81.5 \\
AGTCTGGTCTGGATCTGAGAACTTCAGGCT & 80.6 & 84.7 & 81.9 & 82.2 & 82.5 \\
TCGGAGAAATCACTGAGCTGCCTGAGAAGA & 80.9 & 86.0 & 83.3 & 82.5 & 83.3 \\
CTTCAACGGATCAGGTAGGACTGTGGTGGG & 80.1 & 84.4 & 81.7 & 83.3 & 83.4 \\
ACGCCCACAGGATTAGGCTGGCCCACATTG & 84.0 & 88.9 & 86.2 & 87.5 & 85.5 \\
GTTATTCCGCAGTCCGATGGCAGCAGGCTC & 84.1 & 87.8 & 85.1 & 85.9 & 85.6 \\
TCAGTAGGCGTGACGCAGAGCTGGCGATGG & 84.6 & 88.8 & 86.1 & 88.1 & 88.2 \\
CGCGCCACGTGTGATCTACAGCCGTTCGGC & 84.5 & 88.2 & 85.6 & 89.0 & 89.3 \\
GACCTGACGTGGACCGCTCCTGGGCGTGGT & 86.4 & 89.3 & 86.6 & 91.2 & 89.9 \\
GCCCCTCCACTGGCCGACGGCAGCAGGCTC & 87.7 & 93.3 & 90.6 & 93.8 & 91.5 \\
CGCCGCTGCCGACTGGAGGAGCGCGGGACG & 88.6 & 93.4 & 90.8 & 94.8 & 93.9 \\
\bottomrule
\caption*{\\Melting temperatures of the 92 DNA duplexes studied by Owczarzy \textit{et al.} in Ref.\tcb{$\,^{55}$} at a concentration $c=2\mu$M and 1020mM NaCl. The experimental values ($T^{\rm Exp}$) are compared with predictions obtained with the unzipping parameters by using \tcr{Eq.(10)} ($T_{\rm Bi}^{\rm Unz}$) and \tcr{Eq.(11)} ($T^{\rm Unz}_{\rm Uni}$) for bimolecular and unimolecular reactions, respectively (see main text and \tcr{Sec. 6, Methods}). Finally, $T^{\rm UO}$ and $T^{\rm Hug}$ are obtained with the unified oligonucleotide parameters (UO) in Ref.\tcb{$\,^{43}$} and the Huguet \textit{et al.} (2017) parameters in Ref.\tcb{$\,^{37}$}. Results are reported with errors: $T^{\rm Exp} \pm 1.6^{\circ}$C, $T^{\rm Unz} \pm 1.5^{\circ}$C, $T^{\rm UO} \pm 1.5^{\circ}$C, and $T^{\rm Hug} \pm 1.5^{\circ}$C. Temperatures are given in Celsius degrees.}
\label{tabS:Tmelting}
\end{longtable}
\normalsize
\vfill
\clearpage
\newpage

%NOTE: Ref 55 is owczarcy2008predicting; Ref 43 is santalucia1998unified; Ref 37 is huguet2017derivation

% **********************************************************************
% ---------------------------- SUPP. REFS. -----------------------------
% **********************************************************************

\newpage
\clearpage

\makeatletter
\apptocmd{\thebibliography}{\global\c@NAT@ctr 59\relax}{}{}
\makeatother

\section*{Supplementary References}
\renewcommand{\bibsection}{}

\bibliography{Bibliography}% common bib file